\documentclass[prl,twocolumn,showpacs,nofootinbib,preprintnumbers,amsmath,amssymb,aps,longbibliography,superscriptaddress]{revtex4-2}
\usepackage[T1]{fontenc}
\usepackage[utf8]{inputenc}
\usepackage{subfigure, lmodern, amsmath,amssymb, graphicx, pifont, adjustbox, bm, xcolor}
\usepackage{amsfonts}
\usepackage{amsthm}
\usepackage{comment}
\usepackage{mathtools}
\usepackage{float}
\usepackage{longtable}
\usepackage{graphicx, subfigure}
\usepackage{xcolor}
\usepackage{tikz}
\usepackage{enumitem}
\usepackage{dsfont}
\usepackage{soul}
\usepackage{tikz}
\usepackage[italicdiff]{physics}
\usetikzlibrary{arrows.meta, positioning, decorations.pathreplacing, calc}
\usetikzlibrary{shapes.geometric}

\renewcommand{\imath}{\mathrm{i}}

\def\nn{\nonumber}
\def\Section#1{\noindent {\it #1}}

\makeatletter\g@addto@macro\bfseries{\boldmath}\makeatother

\usepackage[colorlinks=true, citecolor=blue, urlcolor=blue, linkcolor=blue, breaklinks=true, pdfpagelabels=false]{hyperref}

\newcommand{\appendixref}[1]{\hyperref[#1]{appendix~\ref{#1}}}
\def\equationautorefname~#1\null{eq.\,(#1)\null}

\newcommand{\ie}{\begin{equation}\begin{aligned}}
\newcommand{\fe}{\end{aligned}\end{equation}}

\newcommand{\tx}[1]{\text{#1}}

\newcommand{\Breg}{B^{\textrm{reg}}_{L}}
\newcommand{\Birr}{B^{\textrm{irr}}_{L}}
\newcommand{\BregRe}{\bar{B}^{\textrm{reg}}_{L}}
\newcommand{\BirrRe}{\bar{B}^{\textrm{irr}}_{L}}
\newcommand{\ureg}{u^{\textrm{reg}}_\ell}
\newcommand{\uirr}{u^{\textrm{irr}}_\ell}
\newcommand{\Aout}{A^{+}_{L}}
\newcommand{\Ain}{A^{-}_{L}}
\newcommand{\uout}{u^{+}_\ell}
\newcommand{\uin}{u^{-}_\ell}
\newcommand{\QL}{Q}

\newcommand{\Vgrav}{V_{\rm grav}}
\newcommand{\VLove}{V_{\rm Love}}

\providecommand{\Section}[1]{\section{#1}}
\providecommand{\nn}{\nonumber\\}

\usepackage{cleveref}
\crefname{appendix}{App.}{Apps.}
\crefname{equation}{Eq.}{Eqs.}
\crefname{figure}{Fig.}{Figs.}
\crefname{table}{Tab.}{Tabs.}
\crefname{section}{Sec.}{Secs.}
\crefname{definition}{Definition}{Definition}
\crefname{lemma}{Lemma}{Lemma}
\crefname{theorem}{Theorem}{Theorem}
\crefname{proposition}{Proposition}{Proposition}
\crefname{corollary}{Corollary}{Corollary}

\begin{document}

\title{Gravitational Sommerfeld Effects: \\Formalism, Renormalization, and Perturbation to $\mathcal{O}(G^{10})$}
\date{\today}

\author{Chih-Hao Chang}
\affiliation{Department of Physics and Center for Theoretical Physics, National Taiwan University, Taipei 10617, Taiwan}

\author{Chia-Hsien Shen}
\affiliation{Department of Physics and Center for Theoretical Physics, National Taiwan University, Taipei 10617, Taiwan}
\affiliation{Physics Division, National Center for Theoretical Sciences, Taipei 10617, Taiwan}
\affiliation{Max Planck-IAS-NTU Center for Particle Physics, Cosmology and Geometry, Taipei 10617, Taiwan}
\affiliation{Department of Physics and Astronomy, Uppsala University, Box 516, 75120 Uppsala, Sweden}

\author{Zihan Zhou}
\affiliation{Department of Physics, Princeton University, Princeton, New Jersey 08544, USA}

\begin{abstract}
In the effective field theory (EFT) description of binary inspirals, the radiated gravitational waveform receives universal corrections from the curved background, the so-called ``tail effects'', that resum into the so-called ``Sommerfeld factor''. We develop a systematic framework for computing this gravitational Sommerfeld factor for scalar perturbations with the presence of tidal effects on the system. Using the worldline EFT, we recast the diagrammatic resummation as a solution to the $d$-dimensional wave equation with a localized source, and derive a closed-form expression for the Sommerfeld factor in terms of the EFT connection matrix. 
We prove that the phase of the Sommerfeld factor is exactly the same as elastic Compton scattering phase shift when there is no tidal dissipation.
By combining the renormalization techniques in EFT with the Mano--Suzuki--Takasugi method in black hole perturbation theory, we analytically solve the Sommerfeld factor for both the magnitude and phase to $\mathcal{O}(G^{10})$ for the $\ell = 0, 1, 2$ partial waves.
We further establish a new renormalization group equation for the radiative multipole moments, whose exact solution yields an improved resummation of the waveform beyond the universal tail logarithms. These high-precision data and exact relations pave the way for future waveform resummation.
\end{abstract}

\maketitle


\Section{Introduction} ---
The advent of gravitational-wave physics has turned waveform modeling into a precision science. A variety of analytic techniques, including Post-Newtonian (PN) and multipolar Post-Minkowskian (MPM) theory \cite{Blanchet:2013haa,Futamase:2007zz,poisson2014gravity,Blanchet:1985sp,Blanchet:1997jj}, worldline effective field theory (EFT)~\cite{Goldberger:2004jt,Goldberger:2005cd,Porto:2016pyg}, 
the self-force method~\cite{Mano:1996vt,Mano:1996gn,Mino:1996nk,Mino:1997bx,Sasaki:2003xr,Quinn:1996am,Poisson:2011nh,Barack:2018yvs}, 
on-shell  amplitudes~\cite{Cheung:2018wkq,Kosower:2018adc,Bern:2019nnu,Bern:2019crd}, and the effective-one-body (EOB) framework~\cite{Buonanno:1998gg,Buonanno:2000ef,Damour:2001tu}, together with numerical relativity~\cite{Pretorius:2005gq,Baumgarte:2010ndz}, have achieved remarkable success. Yet a systematic physical understanding of the waveform structure, in particular an organizing principle for resummation, remains incomplete.

The waveform from a generic binary system contains a universal sector as a consequence of Birkhoff's theorem: outside of any spherically symmetric source, the gravitational field is described by the Schwarzschild metric regardless of the source's internal structure. As the wave leaves the binary, it propagates under the Schwarzschild background leading to the so-called ``tail'' effects~\cite{DeWitt:1960fc,Blanchet:1987wq,Blanchet:1992br,Blanchet:1993ng}.
This universal behavior includes logarithmic and phase corrections that depend only on the total mass $M$ of the system, but not on its microscopic nature.
From the EFT viewpoint, this universality is made manifest through the renormalization group (RG) running of the radiative multipole moments~\cite{Goldberger:2009qd,Goldberger:2012kf,Almeida:2021jyt}, and it naturally calls for an efficient resummation framework. 
The rest of the waveform is non-universal which depends on system's finite-size structures. 
To study such effects, we need a formalism that incorporates tidal deformations to improve the resummation for the waveform.

The idea of resumming tail effects in the waveform has a long history in the EOB program. Damour and Nagar~\cite{Damour:2007xr,Damour:2007yf} pioneered a ``tail resummation factor'' $T_{\ell m}$ inspired by the Sommerfeld--Coulomb enhancement of quantum scattering, which captures the leading exponentiation of $\pi G M \omega$ and the Coulomb phase shift. This factor has become a standard ingredient in the MPM waveform factorization~\cite{Damour:2008gu,Pan:2011gk,Pompili:2023tna}. Recent progress was made by~\cite{Ivanov:2025ozg}, who derived the universal anomalous dimension of generic multipole moments and proposed an improved resummation using the black hole perturbation theory (BHPT) renormalized angular momentum $\nu$ in place of the integer $\ell$ \cite{Ivanov:2025ozg, Cipriani:2026xmx}, thereby capturing subleading universal logarithms. 
Complementary to this,~\cite{Correia:2024jgr,Caron-Huot:2025tlq} developed a Born-series approach to solve the effective wave equation in worldline EFT, enabling the computation of high-order Compton scattering phase shifts, scalar Love numbers, and the RG running of tidal response functions.

In this Letter, we take a more general and systematic step by computing the \emph{gravitational Sommerfeld factor} for scalar wave perturbations.
In analogous to the Sommerfeld--Coulomb enhancement, we define a the gravitational dressing factor
\ie
\mathcal{S} = \lim_{r\to\infty} \frac{\tx{Waveform}}{\;\;\tx{Waveform} |_{\rm free}} \label{eq: Sommerfeld def} ~,
\fe
where ``free'' means the leading order radiation in a flat spacetime with no tidal effects. 
We will focus on the waveform in the low-frequency regime,
\begin{align}
    GM\omega \ll 1 \,,
\end{align}
such that we can use the worldline EFT to describe the system.
Schematically, the Sommerfeld effect arises from an infinite set of ladder-like diagrams
between the wave emission from a localized source $Q$ and the potential $V$,
\begin{widetext}
\begin{equation}
\begin{aligned} 
\begin{gathered}
    \raisebox{-1.5em}{
\begin{tikzpicture}[x=0.75pt,y=0.75pt,yscale=-1,xscale=1]

\draw [line width=1.5]    (52.2,245.8) -- (135.8,245.8) ;
\draw  [line width=1.0]  (68.34,247.59) .. controls (69.5,247.32) and (70.61,247.06) .. (71.06,246.28) .. controls (71.51,245.49) and (71.19,244.41) .. (70.84,243.26) .. controls (70.5,242.12) and (70.17,241.03) .. (70.62,240.25) .. controls (71.07,239.47) and (72.18,239.21) .. (73.34,238.94) .. controls (74.5,238.66) and (75.61,238.4) .. (76.06,237.62) .. controls (76.51,236.84) and (76.19,235.75) .. (75.84,234.61) .. controls (75.5,233.47) and (75.17,232.38) .. (75.63,231.59) .. controls (76.08,230.81) and (77.18,230.55) .. (78.35,230.28) .. controls (79.51,230.01) and (80.61,229.74) .. (81.06,228.96) .. controls (81.52,228.18) and (81.19,227.09) .. (80.85,225.95) .. controls (80.5,224.81) and (80.18,223.72) .. (80.63,222.93) .. controls (81.08,222.15) and (82.19,221.89) .. (83.35,221.62) .. controls (84.51,221.35) and (85.62,221.09) .. (86.07,220.3) .. controls (86.52,219.52) and (86.19,218.43) .. (85.85,217.29) .. controls (85.5,216.15) and (85.18,215.06) .. (85.63,214.28) .. controls (86.08,213.49) and (87.19,213.23) .. (88.35,212.96) .. controls (89.51,212.69) and (90.62,212.43) .. (91.07,211.64) .. controls (91.52,210.86) and (91.2,209.77) .. (90.85,208.63) .. controls (90.51,207.49) and (90.18,206.4) .. (90.63,205.62) .. controls (91.09,204.83) and (92.19,204.57) .. (93.35,204.3) .. controls (94.51,204.03) and (95.62,203.77) .. (96.07,202.99) .. controls (96.53,202.2) and (96.2,201.11) .. (95.86,199.97) .. controls (95.51,198.83) and (95.18,197.74) .. (95.64,196.96) .. controls (96.09,196.18) and (97.2,195.91) .. (98.36,195.64) .. controls (98.46,195.62) and (98.57,195.59) .. (98.67,195.57) ;
\draw  [color={rgb, 255:red, 0; green, 0; blue, 0 }  ,draw opacity=1 ][fill={rgb, 255:red, 0; green, 0; blue, 0 }  ,fill opacity=1 ] (62.2,246.2) .. controls (62.2,242.67) and (65.07,239.8) .. (68.6,239.8) .. controls (72.13,239.8) and (75,242.67) .. (75,246.2) .. controls (75,249.73) and (72.13,252.6) .. (68.6,252.6) .. controls (65.07,252.6) and (62.2,249.73) .. (62.2,246.2) -- cycle ;
\draw  [line width=1.0]  (102.1,200.94) .. controls (102.93,201.86) and (103.72,202.75) .. (104.63,202.73) .. controls (105.53,202.72) and (106.3,201.81) .. (107.1,200.86) .. controls (107.9,199.9) and (108.66,198.99) .. (109.57,198.98) .. controls (110.47,198.97) and (111.27,199.85) .. (112.1,200.78) .. controls (112.93,201.71) and (113.72,202.59) .. (114.63,202.58) .. controls (115.53,202.56) and (116.3,201.65) .. (117.1,200.7) .. controls (117.9,199.75) and (118.66,198.84) .. (119.57,198.82) .. controls (120.47,198.81) and (121.27,199.69) .. (122.1,200.62) .. controls (122.93,201.55) and (123.72,202.43) .. (124.62,202.42) .. controls (125.53,202.41) and (126.29,201.5) .. (127.1,200.54) .. controls (127.9,199.59) and (128.66,198.68) .. (129.57,198.67) .. controls (130.47,198.65) and (131.26,199.54) .. (132.09,200.46) .. controls (132.92,201.39) and (133.72,202.28) .. (134.62,202.26) .. controls (135.53,202.25) and (136.29,201.34) .. (137.09,200.39) .. controls (137.58,199.81) and (138.05,199.25) .. (138.55,198.89) ;
\begin{scope}[shift={(101.85,221.48)}, xscale=1.5] 
  \draw[fill={rgb,255:red,229; green,229; blue,229}, fill opacity=1, line width=0.75]
    (0,0) ellipse (9.33 and 31.15);
\end{scope}

\node at (101.85,221.48) {$G_R$};

\draw (62.8,254.4) node [anchor=north west][inner sep=0.75pt]  [font=\normalsize]  {$Q$};
\end{tikzpicture}
}
\end{gathered}
& = 
\begin{gathered}
    \raisebox{-4.5em}{
\begin{tikzpicture}[x=0.72pt,y=0.72pt,yscale=-1.2,xscale=1.2]

\draw [line width=1.5]    (112.2,149.2) -- (198.2,149) ;

\draw  [line width=1.0]  (126.43,149.52) .. controls (127.64,149.83) and (128.79,150.12) .. (129.53,149.6) .. controls (130.27,149.08) and (130.4,147.89) .. (130.52,146.65) .. controls (130.65,145.41) and (130.77,144.23) .. (131.51,143.71) .. controls (132.25,143.19) and (133.41,143.48) .. (134.62,143.78) .. controls (135.83,144.09) and (136.98,144.38) .. (137.72,143.86) .. controls (138.46,143.34) and (138.59,142.15) .. (138.71,140.91) .. controls (138.84,139.67) and (138.96,138.49) .. (139.7,137.97) .. controls (140.44,137.45) and (141.6,137.74) .. (142.81,138.04) .. controls (144.02,138.35) and (145.17,138.64) .. (145.91,138.12) .. controls (146.65,137.6) and (146.78,136.42) .. (146.9,135.17) .. controls (147.03,133.93) and (147.15,132.75) .. (147.89,132.23) .. controls (148.63,131.71) and (149.79,132) .. (151,132.31) .. controls (152.21,132.61) and (153.36,132.9) .. (154.1,132.38) .. controls (154.84,131.86) and (154.97,130.68) .. (155.09,129.44) .. controls (155.22,128.19) and (155.34,127.01) .. (156.08,126.49) .. controls (156.82,125.97) and (157.98,126.26) .. (159.19,126.57) .. controls (160.4,126.87) and (161.55,127.16) .. (162.29,126.64) .. controls (163.03,126.12) and (163.15,124.94) .. (163.28,123.7) .. controls (163.41,122.46) and (163.53,121.27) .. (164.27,120.75) .. controls (165.01,120.23) and (166.17,120.52) .. (167.38,120.83) .. controls (168.58,121.13) and (169.74,121.42) .. (170.48,120.9) .. controls (171.22,120.38) and (171.34,119.2) .. (171.47,117.96) .. controls (171.6,116.72) and (171.72,115.53) .. (172.46,115.01) .. controls (173.2,114.5) and (174.36,114.78) .. (175.56,115.09) .. controls (176.77,115.39) and (177.93,115.68) .. (178.67,115.16) .. controls (179.41,114.64) and (179.53,113.46) .. (179.66,112.22) .. controls (179.79,110.98) and (179.91,109.8) .. (180.65,109.28) .. controls (181.39,108.76) and (182.54,109.05) .. (183.75,109.35) .. controls (184.96,109.66) and (186.12,109.94) .. (186.86,109.43) .. controls (187.6,108.91) and (187.72,107.72) .. (187.85,106.48) .. controls (187.98,105.24) and (188.1,104.06) .. (188.84,103.54) .. controls (189.58,103.02) and (190.73,103.31) .. (191.94,103.61) .. controls (193.15,103.92) and (194.31,104.21) .. (195.05,103.69) .. controls (195.61,103.3) and (195.81,102.53) .. (195.94,101.65) ;
\draw  [color={rgb, 255:red, 0; green, 0; blue, 0 }  ,draw opacity=1 ][fill={rgb, 255:red, 0; green, 0; blue, 0 }  ,fill opacity=1 ] (122.2,149.2) .. controls (122.2,145.67) and (125.07,142.8) .. (128.6,142.8) .. controls (132.13,142.8) and (135,145.67) .. (135,149.2) .. controls (135,152.73) and (132.13,155.6) .. (128.6,155.6) .. controls (125.07,155.6) and (122.2,152.73) .. (122.2,149.2) -- cycle ;

\draw (122.8,157.4) node [anchor=north west][inner sep=0.75pt]  [font=\normalsize]  {$Q$};
\end{tikzpicture}
}
\end{gathered}
\quad +
\begin{gathered}      
    \raisebox{-1.5em}{
\begin{tikzpicture}[x=0.75pt,y=0.75pt,yscale=-1,xscale=1]

\draw [line width=1.5]    (52.2,245.8) -- (135.8,245.8) ;
\draw  [line width=1.0]  (68.34,247.59) .. controls (69.5,247.32) and (70.61,247.06) .. (71.06,246.28) .. controls (71.51,245.49) and (71.19,244.41) .. (70.84,243.26) .. controls (70.5,242.12) and (70.17,241.03) .. (70.62,240.25) .. controls (71.07,239.47) and (72.18,239.21) .. (73.34,238.94) .. controls (74.5,238.66) and (75.61,238.4) .. (76.06,237.62) .. controls (76.51,236.84) and (76.19,235.75) .. (75.84,234.61) .. controls (75.5,233.47) and (75.17,232.38) .. (75.63,231.59) .. controls (76.08,230.81) and (77.18,230.55) .. (78.35,230.28) .. controls (79.51,230.01) and (80.61,229.74) .. (81.06,228.96) .. controls (81.52,228.18) and (81.19,227.09) .. (80.85,225.95) .. controls (80.5,224.81) and (80.18,223.72) .. (80.63,222.93) .. controls (81.08,222.15) and (82.19,221.89) .. (83.35,221.62) .. controls (84.51,221.35) and (85.62,221.09) .. (86.07,220.3) .. controls (86.52,219.52) and (86.19,218.43) .. (85.85,217.29) .. controls (85.5,216.15) and (85.18,215.06) .. (85.63,214.28) .. controls (86.08,213.49) and (87.19,213.23) .. (88.35,212.96) .. controls (89.51,212.69) and (90.62,212.43) .. (91.07,211.64) .. controls (91.52,210.86) and (91.2,209.77) .. (90.85,208.63) .. controls (90.51,207.49) and (90.18,206.4) .. (90.63,205.62) .. controls (91.09,204.83) and (92.19,204.57) .. (93.35,204.3) .. controls (94.51,204.03) and (95.62,203.77) .. (96.07,202.99) .. controls (96.53,202.2) and (96.2,201.11) .. (95.86,199.97) .. controls (95.51,198.83) and (95.18,197.74) .. (95.64,196.96) .. controls (96.09,196.18) and (97.2,195.91) .. (98.36,195.64) .. controls (98.46,195.62) and (98.57,195.59) .. (98.67,195.57) ;
\draw[fill=black] (68.6,245.8) circle (6.4);
\draw  [line width=1.0]  (102.1,200.94) .. controls (102.93,201.86) and (103.72,202.75) .. (104.63,202.73) .. controls (105.53,202.72) and (106.3,201.81) .. (107.1,200.86) .. controls (107.9,199.9) and (108.66,198.99) .. (109.57,198.98) .. controls (110.47,198.97) and (111.27,199.85) .. (112.1,200.78) .. controls (112.93,201.71) and (113.72,202.59) .. (114.63,202.58) .. controls (115.53,202.56) and (116.3,201.65) .. (117.1,200.7) .. controls (117.9,199.75) and (118.66,198.84) .. (119.57,198.82) .. controls (120.47,198.81) and (121.27,199.69) .. (122.1,200.62) .. controls (122.93,201.55) and (123.72,202.43) .. (124.62,202.42) .. controls (125.53,202.41) and (126.29,201.5) .. (127.1,200.54) .. controls (127.9,199.59) and (128.66,198.68) .. (129.57,198.67) .. controls (130.47,198.65) and (131.26,199.54) .. (132.09,200.46) .. controls (132.92,201.39) and (133.72,202.28) .. (134.62,202.26) .. controls (135.53,202.25) and (136.29,201.34) .. (137.09,200.39) .. controls (137.58,199.81) and (138.05,199.25) .. (138.55,198.89) ;
\draw[fill={rgb,255:red,229;green,229;blue,229}, line width=0.75]
(102,220.2) ellipse (9 and 31);

\draw (95.2,215.2) node [anchor=north west][inner sep=0.75pt]  [font=\normalsize]  {$V$};
\draw (62.8,254.4) node [anchor=north west][inner sep=0.75pt]  [font=\normalsize]  {$Q$};
\end{tikzpicture}
}
\end{gathered} \quad +
\begin{gathered}    
    \raisebox{-1.5em}{
\begin{tikzpicture}[x=0.75pt,y=0.75pt,yscale=-1,xscale=1]

\draw [line width=1.5]    (52.2,246.2) -- (174.2,245.8) ;
\draw  [line width=1.0]  (68.34,247.59) .. controls (69.5,247.32) and (70.61,247.06) .. (71.06,246.28) .. controls (71.51,245.49) and (71.19,244.41) .. (70.84,243.26) .. controls (70.5,242.12) and (70.17,241.03) .. (70.62,240.25) .. controls (71.07,239.47) and (72.18,239.21) .. (73.34,238.94) .. controls (74.5,238.66) and (75.61,238.4) .. (76.06,237.62) .. controls (76.51,236.84) and (76.19,235.75) .. (75.84,234.61) .. controls (75.5,233.47) and (75.17,232.38) .. (75.63,231.59) .. controls (76.08,230.81) and (77.18,230.55) .. (78.35,230.28) .. controls (79.51,230.01) and (80.61,229.74) .. (81.06,228.96) .. controls (81.52,228.18) and (81.19,227.09) .. (80.85,225.95) .. controls (80.5,224.81) and (80.18,223.72) .. (80.63,222.93) .. controls (81.08,222.15) and (82.19,221.89) .. (83.35,221.62) .. controls (84.51,221.35) and (85.62,221.09) .. (86.07,220.3) .. controls (86.52,219.52) and (86.19,218.43) .. (85.85,217.29) .. controls (85.5,216.15) and (85.18,215.06) .. (85.63,214.28) .. controls (86.08,213.49) and (87.19,213.23) .. (88.35,212.96) .. controls (89.51,212.69) and (90.62,212.43) .. (91.07,211.64) .. controls (91.52,210.86) and (91.2,209.77) .. (90.85,208.63) .. controls (90.51,207.49) and (90.18,206.4) .. (90.63,205.62) .. controls (91.09,204.83) and (92.19,204.57) .. (93.35,204.3) .. controls (94.51,204.03) and (95.62,203.77) .. (96.07,202.99) .. controls (96.53,202.2) and (96.2,201.11) .. (95.86,199.97) .. controls (95.51,198.83) and (95.18,197.74) .. (95.64,196.96) .. controls (96.09,196.18) and (97.2,195.91) .. (98.36,195.64) .. controls (98.46,195.62) and (98.57,195.59) .. (98.67,195.57) ;
\draw  [color={rgb, 255:red, 0; green, 0; blue, 0 }  ,draw opacity=1 ][fill={rgb, 255:red, 0; green, 0; blue, 0 }  ,fill opacity=1 ] (62.2,246.2) .. controls (62.2,242.67) and (65.07,239.8) .. (68.6,239.8) .. controls (72.13,239.8) and (75,242.67) .. (75,246.2) .. controls (75,249.73) and (72.13,252.6) .. (68.6,252.6) .. controls (65.07,252.6) and (62.2,249.73) .. (62.2,246.2) -- cycle ;
\draw  [line width=1.0]  (102.1,200.42) .. controls (102.93,201.29) and (103.72,202.11) .. (104.62,202.1) .. controls (105.53,202.09) and (106.29,201.23) .. (107.1,200.34) .. controls (107.9,199.45) and (108.66,198.59) .. (109.57,198.58) .. controls (110.47,198.57) and (111.27,199.39) .. (112.1,200.26) .. controls (112.92,201.13) and (113.72,201.96) .. (114.62,201.94) .. controls (115.53,201.93) and (116.29,201.08) .. (117.09,200.18) .. controls (117.9,199.29) and (118.66,198.44) .. (119.57,198.42) .. controls (120.47,198.41) and (121.26,199.24) .. (122.09,200.1) .. controls (122.92,200.97) and (123.72,201.8) .. (124.62,201.79) .. controls (125.53,201.77) and (126.29,200.92) .. (127.09,200.03) .. controls (127.89,199.13) and (128.66,198.28) .. (129.57,198.27) .. controls (130.47,198.25) and (131.26,199.08) .. (132.09,199.95) .. controls (132.92,200.82) and (133.71,201.64) .. (134.62,201.63) .. controls (135.52,201.61) and (136.29,200.76) .. (137.09,199.87) .. controls (137.89,198.97) and (138.66,198.12) .. (139.56,198.11) .. controls (140.47,198.09) and (141.26,198.92) .. (142.09,199.79) .. controls (142.92,200.66) and (143.71,201.49) .. (144.62,201.47) .. controls (145.52,201.46) and (146.29,200.61) .. (147.09,199.71) .. controls (147.89,198.82) and (148.66,197.97) .. (149.56,197.95) .. controls (150.47,197.94) and (151.26,198.76) .. (152.09,199.63) .. controls (152.92,200.5) and (153.71,201.33) .. (154.62,201.31) .. controls (155.52,201.3) and (156.29,200.45) .. (157.09,199.55) .. controls (157.89,198.66) and (158.66,197.81) .. (159.56,197.79) .. controls (160.47,197.78) and (161.26,198.61) .. (162.09,199.48) .. controls (162.92,200.34) and (163.71,201.17) .. (164.62,201.16) .. controls (165.52,201.14) and (166.29,200.29) .. (167.09,199.4) .. controls (167.89,198.5) and (168.66,197.65) .. (169.56,197.64) .. controls (170.47,197.62) and (171.26,198.45) .. (172.09,199.32) .. controls (172.91,200.18) and (173.7,201.01) .. (174.6,201) ;

\draw[fill={rgb,255:red,229;green,229;blue,229}, line width=0.75]
(101.0,220.2) ellipse (9 and 31);

\draw[fill={rgb,255:red,229;green,229;blue,229}, line width=0.75]
(142.2,220.2) ellipse (9 and 31);

\draw (94.2,214.2) node [anchor=north west][inner sep=0.75pt]  [font=\normalsize]  {$V$};
\draw (62.8,254.4) node [anchor=north west][inner sep=0.75pt]  [font=\normalsize]  {$Q$};
\draw (135.4,214.2) node [anchor=north west][inner sep=0.75pt]  [font=\normalsize]  {$V$};
\end{tikzpicture}
}
\end{gathered}
\end{aligned} + \cdots
\end{equation} 
\end{widetext}
where $G_R$ denotes the full retarded Green's function. Here, the potential $V$ includes the two-particle irreducible diagrams in a perturbative series in $G$. More precisely, $V$ includes both the long-distance gravitational potential and the short-distance contact interactions from tidal effects.
The corresponding resummation is naturally organized by the Lippmann--Schwinger
equation,
\begin{equation}
\label{eq:Lip_Sch}
\begin{aligned}
    \phi & = \phi^{(0)} + \phi^{(0)} V G_R^{(0)} + \phi^{(0)} V G_R^{(0)} V G_R^{(0)} + \cdots \\
    & = \phi^{(0)} + \phi V G_R^{(0)} \,,
\end{aligned}
\end{equation}
where $G_R^{(0)}$ is the free retarded Green's function. This equation is exactly the integral form of the scalar Teukolsky equation with a localized source term
\begin{equation}
    (\nabla^2 + \omega^2) \phi = V \phi + Q \delta^{(3)}(x).
\end{equation}
In this language, the Sommerfeld enhancement factor defined in Eq.~\eqref{eq: Sommerfeld def} can be written as
\begin{equation}
\begin{aligned}
    \mathcal{S} & = \lim_{r\to\infty} \frac{\phi}{\phi|_{\rm free}}  = 
    \begin{gathered}
        \scalebox{0.7}{\raisebox{-1.5em}{
\begin{tikzpicture}[x=0.75pt,y=0.75pt,yscale=-1,xscale=1]

\draw [line width=1.5]    (52.2,245.8) -- (135.8,245.8) ;
\draw  [line width=1.0]  (68.34,247.59) .. controls (69.5,247.32) and (70.61,247.06) .. (71.06,246.28) .. controls (71.51,245.49) and (71.19,244.41) .. (70.84,243.26) .. controls (70.5,242.12) and (70.17,241.03) .. (70.62,240.25) .. controls (71.07,239.47) and (72.18,239.21) .. (73.34,238.94) .. controls (74.5,238.66) and (75.61,238.4) .. (76.06,237.62) .. controls (76.51,236.84) and (76.19,235.75) .. (75.84,234.61) .. controls (75.5,233.47) and (75.17,232.38) .. (75.63,231.59) .. controls (76.08,230.81) and (77.18,230.55) .. (78.35,230.28) .. controls (79.51,230.01) and (80.61,229.74) .. (81.06,228.96) .. controls (81.52,228.18) and (81.19,227.09) .. (80.85,225.95) .. controls (80.5,224.81) and (80.18,223.72) .. (80.63,222.93) .. controls (81.08,222.15) and (82.19,221.89) .. (83.35,221.62) .. controls (84.51,221.35) and (85.62,221.09) .. (86.07,220.3) .. controls (86.52,219.52) and (86.19,218.43) .. (85.85,217.29) .. controls (85.5,216.15) and (85.18,215.06) .. (85.63,214.28) .. controls (86.08,213.49) and (87.19,213.23) .. (88.35,212.96) .. controls (89.51,212.69) and (90.62,212.43) .. (91.07,211.64) .. controls (91.52,210.86) and (91.2,209.77) .. (90.85,208.63) .. controls (90.51,207.49) and (90.18,206.4) .. (90.63,205.62) .. controls (91.09,204.83) and (92.19,204.57) .. (93.35,204.3) .. controls (94.51,204.03) and (95.62,203.77) .. (96.07,202.99) .. controls (96.53,202.2) and (96.2,201.11) .. (95.86,199.97) .. controls (95.51,198.83) and (95.18,197.74) .. (95.64,196.96) .. controls (96.09,196.18) and (97.2,195.91) .. (98.36,195.64) .. controls (98.46,195.62) and (98.57,195.59) .. (98.67,195.57) ;
\draw  [color={rgb, 255:red, 0; green, 0; blue, 0 }  ,draw opacity=1 ][fill={rgb, 255:red, 0; green, 0; blue, 0 }  ,fill opacity=1 ] (62.2,246.2) .. controls (62.2,242.67) and (65.07,239.8) .. (68.6,239.8) .. controls (72.13,239.8) and (75,242.67) .. (75,246.2) .. controls (75,249.73) and (72.13,252.6) .. (68.6,252.6) .. controls (65.07,252.6) and (62.2,249.73) .. (62.2,246.2) -- cycle ;
\draw  [line width=1.0]  (102.1,200.94) .. controls (102.93,201.86) and (103.72,202.75) .. (104.63,202.73) .. controls (105.53,202.72) and (106.3,201.81) .. (107.1,200.86) .. controls (107.9,199.9) and (108.66,198.99) .. (109.57,198.98) .. controls (110.47,198.97) and (111.27,199.85) .. (112.1,200.78) .. controls (112.93,201.71) and (113.72,202.59) .. (114.63,202.58) .. controls (115.53,202.56) and (116.3,201.65) .. (117.1,200.7) .. controls (117.9,199.75) and (118.66,198.84) .. (119.57,198.82) .. controls (120.47,198.81) and (121.27,199.69) .. (122.1,200.62) .. controls (122.93,201.55) and (123.72,202.43) .. (124.62,202.42) .. controls (125.53,202.41) and (126.29,201.5) .. (127.1,200.54) .. controls (127.9,199.59) and (128.66,198.68) .. (129.57,198.67) .. controls (130.47,198.65) and (131.26,199.54) .. (132.09,200.46) .. controls (132.92,201.39) and (133.72,202.28) .. (134.62,202.26) .. controls (135.53,202.25) and (136.29,201.34) .. (137.09,200.39) .. controls (137.58,199.81) and (138.05,199.25) .. (138.55,198.89) ;
\begin{scope}[shift={(101.85,221.48)}, xscale=1.5] 
  \draw[fill={rgb,255:red,229; green,229; blue,229}, fill opacity=1, line width=0.75]
    (0,0) ellipse (9.33 and 31.15);
\end{scope}

\node at (101.85,221.48) {$G_R$};

\draw (62.8,254.4) node [anchor=north west][inner sep=0.75pt]  [font=\normalsize]  {$Q$};
\end{tikzpicture}
}}
    \end{gathered}
     \Bigg/
    \begin{gathered}
        \scalebox{0.7}{\raisebox{-4.5em}{
\begin{tikzpicture}[x=0.72pt,y=0.72pt,yscale=-1.2,xscale=1.2]

\draw [line width=1.5]    (112.2,149.2) -- (198.2,149) ;

\draw  [line width=1.0]  (126.43,149.52) .. controls (127.64,149.83) and (128.79,150.12) .. (129.53,149.6) .. controls (130.27,149.08) and (130.4,147.89) .. (130.52,146.65) .. controls (130.65,145.41) and (130.77,144.23) .. (131.51,143.71) .. controls (132.25,143.19) and (133.41,143.48) .. (134.62,143.78) .. controls (135.83,144.09) and (136.98,144.38) .. (137.72,143.86) .. controls (138.46,143.34) and (138.59,142.15) .. (138.71,140.91) .. controls (138.84,139.67) and (138.96,138.49) .. (139.7,137.97) .. controls (140.44,137.45) and (141.6,137.74) .. (142.81,138.04) .. controls (144.02,138.35) and (145.17,138.64) .. (145.91,138.12) .. controls (146.65,137.6) and (146.78,136.42) .. (146.9,135.17) .. controls (147.03,133.93) and (147.15,132.75) .. (147.89,132.23) .. controls (148.63,131.71) and (149.79,132) .. (151,132.31) .. controls (152.21,132.61) and (153.36,132.9) .. (154.1,132.38) .. controls (154.84,131.86) and (154.97,130.68) .. (155.09,129.44) .. controls (155.22,128.19) and (155.34,127.01) .. (156.08,126.49) .. controls (156.82,125.97) and (157.98,126.26) .. (159.19,126.57) .. controls (160.4,126.87) and (161.55,127.16) .. (162.29,126.64) .. controls (163.03,126.12) and (163.15,124.94) .. (163.28,123.7) .. controls (163.41,122.46) and (163.53,121.27) .. (164.27,120.75) .. controls (165.01,120.23) and (166.17,120.52) .. (167.38,120.83) .. controls (168.58,121.13) and (169.74,121.42) .. (170.48,120.9) .. controls (171.22,120.38) and (171.34,119.2) .. (171.47,117.96) .. controls (171.6,116.72) and (171.72,115.53) .. (172.46,115.01) .. controls (173.2,114.5) and (174.36,114.78) .. (175.56,115.09) .. controls (176.77,115.39) and (177.93,115.68) .. (178.67,115.16) .. controls (179.41,114.64) and (179.53,113.46) .. (179.66,112.22) .. controls (179.79,110.98) and (179.91,109.8) .. (180.65,109.28) .. controls (181.39,108.76) and (182.54,109.05) .. (183.75,109.35) .. controls (184.96,109.66) and (186.12,109.94) .. (186.86,109.43) .. controls (187.6,108.91) and (187.72,107.72) .. (187.85,106.48) .. controls (187.98,105.24) and (188.1,104.06) .. (188.84,103.54) .. controls (189.58,103.02) and (190.73,103.31) .. (191.94,103.61) .. controls (193.15,103.92) and (194.31,104.21) .. (195.05,103.69) .. controls (195.61,103.3) and (195.81,102.53) .. (195.94,101.65) ;
\draw  [color={rgb, 255:red, 0; green, 0; blue, 0 }  ,draw opacity=1 ][fill={rgb, 255:red, 0; green, 0; blue, 0 }  ,fill opacity=1 ] (122.2,149.2) .. controls (122.2,145.67) and (125.07,142.8) .. (128.6,142.8) .. controls (132.13,142.8) and (135,145.67) .. (135,149.2) .. controls (135,152.73) and (132.13,155.6) .. (128.6,155.6) .. controls (125.07,155.6) and (122.2,152.73) .. (122.2,149.2) -- cycle ;

\draw (122.8,157.4) node [anchor=north west][inner sep=0.75pt]  [font=\normalsize]  {$Q$};
\end{tikzpicture}
}}
    \end{gathered}
     .
\end{aligned}
\end{equation}

Beyond this conceptual improvement, we develop an efficient computational formalism that leverages the EFT Born series~\cite{Correia:2024jgr,Caron-Huot:2025tlq} in the near zone while using the Mano–Suzuki–Takasugi (MST) method in BHPT~\cite{Mano:1996vt,Mano:1996gn,Mino:1996nk,Mino:1997bx,Sasaki:2003xr} to extend the EFT results to the far zone.
Using this hybrid approach, we are able to compute the Sommerfeld factor, for both its magnitude and phase, perturbatively to $\mathcal{O}(G^{10})$ for the $\ell = 0, 1, 2$ partial waves~\cite{Sommerfeld_data}. This provides a clean path for future extension to the spin-2 case.
See~\cite{Fucito:2024wlg,Cipriani:2025ikx,Cipriani:2026xmx,Cipriani:2026myb,Combaluzier--Szteinsznaider:2025eoc,Kobayashi:2025vgl,Kosmopoulos:2025rfj} for other studies in this direction.

Using these high-order data, we examine the existing resummation proposals. 
As is well-known, the Damour--Nagar formula captures the leading Coulomb structure but misses corrections starting at $\mathcal{O}(G^2)$, while the improved resummation of~\cite{Ivanov:2025ozg,Cipriani:2026xmx} correctly captures all universal logarithms through $\mathcal{O}(G^{2\ell+1})$. We further propose a new resummation that uses the full EFT anomalous dimension including the mixing between radiative multipoles and tidal response function, which goes beyond the universal sector and provides an improved match to the exact Sommerfeld factor at higher orders.

\Section{The Wave Equation with a Source.} ---
We will study the waveform of a scalar $\phi$ emitted from a spinless compact system with mass $M$, which creates a $d$-dimensional Schwarzschild background
\begin{align}
    ds^2 &= -f dt^2 +f^{-1} dr^2 + r^2 d\Omega_{d-2}^2,
    \label{eq:Schwarzschild}
\end{align}
where
\begin{align}
    f &= 1- \frac{2GM n_d \mu^{4-d}}{r^{d-3}}, \quad
    n_d \equiv \frac{4\pi^{\frac{3-d}{2}} \Gamma\left(\frac{d-1}{2}\right)}{d-2}\,.
\end{align}
We have used dimensional regularization to promote $d=4-2\epsilon$ and introduced $\mu$ as the renormalization scale.
The source for $\phi$ is encoded by a series of multipole moments $Q^L(\tau)$ determined by the short-distance dynamics of the compact system. The action reads
\ie
S_\tx{source} =  \sum_\ell \mu^{\frac{4-d}{2}} \int \dd{\tau}  Q^L(\tau) \nabla_L \phi \,,
\fe
where we use $L$ to denote a collection of $\ell$ Lorentz indices. The derivative $\nabla_L$ is a $\ell$-th order spatial derivative projected to the symmetric and trace-free (STF) part in the rest frame of the worldline.
The full action is then
\ie
S &= S_\tx{bulk} + S_\tx{Love} + S_\tx{source},
\fe
where $S_\tx{bulk}$ describes $\phi$ minimally coupled to gravity and $S_\tx{Love}$ introduces the tidal interaction on the worldline
\ie
S_\tx{Love} = \sum_{n,\ell} \frac{ \mu^{4-d} }{\ell!} C_{\ell,n} \int \dd{\tau} \nabla_L \phi_+ \nabla^L \partial^n_\tau\phi_-,
\label{eq:action_Love}
\fe
with $C_{\ell,n}$ being the Love number. 
To account for dissipative tidal effects described by odd in $n$ terms in Eq.~\eqref{eq:action_Love}, 
we use the in-in formalism to split the field as $\phi_{\pm}$ and the equation of motion for the physical $\phi_+$ is given by $\delta S/\delta \phi_- =0$. We will drop the $+$ subscript from now on.
The wave equation is then a Klein-Gordon equation of $\phi$ in the Schwarzschild spacetime with tidal interaction and sources on the worldline.
We render the complete action and the wave equation in the Supplemental Material.

Due to the rotational symmetry, we can decompose the scalar field as
\begin{align}
\phi (t,\vb*{r})
=& \sum_{L} e^{-i\omega t} R_{L} (r) Y_{L}(\hat{\vb*{r}}) 
\label{eq: scalar wave spherical decomposition}
\end{align}
where $Y_{L}$ is the $d$-dimensional spherical harmonics in the STF basis.
Defining $R_{L} = u_{L}/r^{1-\epsilon}$, the wave equation becomes one dimensional
\ie
\left( \dv[2]{}{r} + \omega^2 \!-\! \frac{(\ell - \epsilon)(\ell - \epsilon + 1)}{r^2} -\! V(r) \!\right) u_{L} = \rho_{L} \,,
\label{eq: radial equation with source}
\fe
where the potential $V(r)=\Vgrav(r)+\VLove(r)$ is sum of the long-range  $\Vgrav(r)$ and the short-range $\VLove(r)$ from tidal interaction that depends on the Love number through the tidal response function $F_\ell \equiv \sum_{n} C_{\ell,n} (i\omega)^n$.
The new ingredient compared to~\cite{Caron-Huot:2025tlq} is the source term
\ie 
\rho_{L} = \frac{c_\ell}{\ell!}(-1)^{\ell+1} \mu^{\epsilon} r^{1-\epsilon} Q^{L'}(\omega) \!\int_\Omega Y^{*}_{L} \nabla_{L'} \delta(\vb*{r}) \label{eq: def of source}
\fe
where $\QL^{L'}(\omega)$ is the multipole moment in the frequency space and $\int_\Omega$ integrates over the $d-2$ sphere. We will drop the argument of $\QL^{L'}$ when there is no confusion.
We put the details on wave equation and spherical harmonics in the Supplemental Material.

\Section{Derivation of the Sommerfeld Factor} ---
In the presence of the source, one can solve the waveform using Green's function. But to recycle the known homogeneous solutions, we observe that the source is localized at $r=0$; therefore, the waveform are still given by the same homogeneous solutions. The only role of this source is to modify the boundary conditions at $r=0$. Both the Green's function and the modified boundary conditions yield the same result. We only present the derivation from the latter and see more details in the Supplemental Material.

As explained in~\cite{Caron-Huot:2025tlq}, we can span the waveform in terms of the two independent solutions, $\ureg$ and $\uirr$,
\begin{align}
    u_{L} =&  \Breg \ureg + \Birr \uirr \label{eq:wave_NZ} \\
    \rightarrow & \mu^{-\epsilon} \Breg \, r^{\ell+1-\epsilon}
    +\frac{ \mu^{\epsilon}}{2\ell+1-2\epsilon} \Birr \,r^{-\ell+\epsilon} +\dots \,,\nn
\end{align}
where $\Breg$ and $\Birr$ are two unknown coefficients, and we show the near-zone (NZ) behavior near $r\rightarrow 0$ in the second line\footnote{We choose a different $\mu$ dependence from \cite{Caron-Huot:2025tlq} but it does not affect the Sommerfeld factors and renormalization group equations.}.
A crucial insight from~\cite{Caron-Huot:2025tlq} is that the tidal potential $V_{\rm Love}(r)$ provides a boundary condition on $\Breg$ and $\Birr$. This is relatively simple to see for $\ell=0$ since
\begin{align}
    \nabla^2 \frac{r^{\epsilon}}{r^{1-\epsilon}} = -\frac{4\pi^{ \frac{3}{2}-\epsilon} }{\Gamma \left( \frac{1}{2}-\epsilon \right) } \delta^{3-2\epsilon}(\vb*{r}) \,,
\end{align}
such that the $\nabla^2$ acting on $\uirr$ yields a delta function that must be matched to the one in $\VLove(r)$.
For our case, the source itself is also an $\ell$-th derivative on the delta function. Therefore, the presence of the source modifies the NZ boundary condition as
\begin{align}
\frac{\Birr}{c_\ell} - F_{\ell} \Breg = \QL_L \,,
\,
c_\ell = \frac{2^{\ell}\,\Gamma\left(\frac{3}{2}-\epsilon + \ell \right)}{2\pi^{\frac{3}{2}-\epsilon}}
\label{eq: near zone boundary condition}
\end{align}
whose detailed derivation is given in the Supplemental Material. The boundary condition in~\cite{Caron-Huot:2025tlq} is then the source-free limit of Eq.~\eqref{eq: near zone boundary condition}.

Once we fix the NZ boundary condition, the waveform is then fixed by imposing another boundary condition in the far zone (FZ), $\omega r\gg 1$. 
We choose the two independent solutions as the outgoing  $\uout$ and the incoming $\uin$, which satisfy large-$r$ behavior
\ie
u^{\pm}_\ell (r) \xrightarrow{r \to \infty} (\mp i)^{\ell + 1}  e^{\pm i \omega \big(r + 2GM \log(\bar\mu_{\rm IR} r) \big) },
\label{eq: EFT far zone asymptotic bc}
\fe
where $\bar \mu_{\rm IR}$ is the IR cutoff originating from the $1/r$ potential.
The full waveform is
\begin{align}
    u_L = \Aout \uout
    + \Ain \uin \,.
    \label{eq:wave_FZ}
\end{align}
For the waveform emission, we assume that there is no incoming wave, and thus we impose $\Ain = 0$.

To combine the boundary conditions in the NZ and FZ,
we note that the wave equation is of second order. The bases in far and near zones must be linearly related,  $(\ureg,\uirr)= (\uout,\uin) \,\vb*{W}$, where $\vb*{W}$ is the $2\times 2$ \emph{connection matrix} that only depends the long-range $\Vgrav$. Since $\Vgrav$ is real, the two solutions $(\uout,\uin)$ must be related by complex conjugation, implying $W_{2i}=W_{1i}^*$.
By equating Eqs.~\eqref{eq:wave_NZ} and \eqref{eq:wave_FZ}, we find
\ie
\begin{pmatrix}
\Aout \\ \Ain
\end{pmatrix} = 
\vb*{W} \begin{pmatrix}
\Breg \\[3pt]
\Birr
\end{pmatrix} \,.
\label{eq: EFT connection matrix def}
\fe
Solving $\Aout$ from Eqs.~\eqref{eq: near zone boundary condition},\,\eqref{eq: EFT connection matrix def}, and $\Ain=0$, we arrive at the main formula of our paper
\begin{align}
\phi
\xrightarrow{r\to\infty}
& \sum_{\ell} \mathcal{S}_\ell \frac{ i c_\ell\,\omega^{\ell}}{(2\ell+1)!!}
\, \frac{e^{-i\omega t}u^+_\ell}{r^{1-\epsilon}} Q^L Y_L(\hat{\vb*{r}}) \nn \\
\mathcal{S}_\ell
=& \frac{ (2\ell+1) !!}{2\omega^{\ell+1}(W_{21} + c_\ell F_\ell W_{22})}
\label{eq:phi_final}
\end{align}
where $\mathcal{S}_\ell$ is the Sommerfeld factor in Eq.~\eqref{eq: Sommerfeld def} that is normalized to unity in the free theory limit in $d = 4$.
At this stage, the bare connection matrix $W_{ij}$ may be UV divergent and therefore requires renormalization, which we will discuss later.

One can compare the Sommerfeld effect with the Compton scattering, in which there is an incoming wave but no source. 
The corresponding $S$-matrix $\hat{S}_\ell \equiv \Aout/\Ain$ is given by~\cite{Caron-Huot:2025tlq}
\begin{align}
\hat{S}_\ell =& \frac{\Aout}{\Ain} = \frac{W_{11}+c_\ell F_\ell W_{12}}{W_{21}+c_\ell F_\ell W_{22}}
\label{eq:S_matrix}
\end{align}
An important case is the conservative system where all tidal interactions have even numbers of time derivatives, which leads to a real $F_\ell$.
Combining this conservative condition with $W_{2i}=W_{1i}^*$, we find
\begin{align}
    \textrm{Arg} \big(\mathcal{S}_\ell \big)= \frac{1}{2}\,\textrm{Arg}\big(\hat{S}_\ell \big)\,.
\label{eq:Sommerfeld_phase_Compton_phase}
\end{align}
This is an exact relation between the phase in the Sommerfeld factor and the phase shift in the conservative Compton scattering.
Through the EFT setup, this relation allows us to import the rich results in Compton scattering to the waveform emission~\cite{Caron-Huot:2025tlq,Bautista:2026qse,Bjerrum-Bohr:2026fhs,Ivanov:2026icp}.

\Section{Near Zone and Renormalization} ---
Armed with the general formula for the Sommerfeld factor, the task is to compute the connection matrix. As laid out in~\cite{Caron-Huot:2025tlq}, one can calculate this using the Born series. This is particularly straightforward in the NZ since the perturbative expansion of $\ureg$ and $\uirr$ are polynomials in $r$ which are simple to integrate in general $d$.

However, the point-particle limit we take in the EFT leads to ultraviolet (UV) divergence at $r=0$ at $d=4$.
Our formula~\eqref{eq:phi_final} is at the bare level. The physical waveform requires a renormalization procedure. 
We start the renormalization as in~\cite{Caron-Huot:2025tlq} by constructing the renormalized NZ coefficients as $(\BregRe,\BirrRe)^T = Z(\Breg, \Birr)^T$. Demanding the renormalization-scale independence of the bare parameters then gives $(d/{d\log\mu}  +\vb*{\gamma} ) (\bar{B}_\ell^\tx{reg}, \bar{B}_\ell^\tx{irr})^T= 0$, where the anomalous dimension $\gamma$ is reviewed in Eq.~\eqref{eq:B_RG}.
Using the renormalized boundary condition~\eqref{eq: near zone boundary condition}, we find
\begin{align}
    \frac{d}{d\log\mu} \bar{F}_\ell(\omega) &= 
    - \frac{\gamma_{21}}{c_\ell} + (\gamma_{11} - \gamma_{22}) \bar{F}_\ell(\omega) + c_\ell \gamma_{12} \bar{F}^2_\ell(\omega) \nn
    \\
    \frac{d}{d\log\mu} \bar{Q}_L(\omega) &= \Big( \gamma_{11}  + \gamma_{12} c_\ell \bar{F}_\ell (\omega) \Big) \bar{Q}_L (\omega)\,, 
    \label{eq: RG of multipole moments}
\end{align}
where the second equation is new: it gives the full RG equation of linearized source with the full tidal-response dependence.  
We will discuss the structures and analytic solutions to the RG equations later.

\Section{Perturbative data from BHPT} ---
After computing the NZ solutions and renormalization, the remaining task is to calculate the FZ solutions. 
While this can be achieved using the Born series, the integration becomes more involved~\cite{Caron-Huot:2025tlq}. In this paper, we instead aim to recycle techniques from BHPT to bypass nontrivial integration.

\begin{figure}[t]
    \scalebox{0.8}{
        \begin{tikzpicture}[
    node distance = 2.6cm,
    every node/.style = {font=\large},
    zone/.style = {font=\Large\bfseries},
    arrowR/.style = {->, very thick, black!70},
    arrowB/.style = {->, very thick, black}]
\node (Rreg) {$R_\ell^\tx{reg}, R_\ell^\tx{irr}$};

\node[above = of Rreg] (Rc) {$R_\nu^C, R_{-\nu-1}^C$};

\node[zone, below = 0.4cm of Rreg] {Near Zone};

\node[right = 2.7 cm of Rreg] (Routin) {$R_\ell^\tx{+}, R_\ell^\tx{-}$};

\node[above=of Routin] (Rupdown) {$R_\nu^\tx{up}, R_\nu^\tx{down}$};

\node[zone, below = 0.4cm of Routin] {Far Zone};


\draw[arrowR] (Rreg) -- node[left=6pt] {$\vb*{T}^\tx{NZ}$} (Rc);

\draw[arrowR] (Rupdown) -- node[right=6pt] {$\vb*{T}^\tx{FZ}$} (Routin);

\draw[arrowR] (Rc) -- node[above=6pt] {$\vb*{W}^{\text{BHPT}}$} (Rupdown);

\draw[black, very thick,->] (0.5,0.5) .. controls (0.5,4) and (4.0,4) .. (4.0 ,0.5) node[midway, below=8pt, font=\Large] {${\vb*{W}}$};

\node[right=0.4 cm of Rupdown] { \textbf{BHPT} };
\node[right=0.6 cm of Routin] {\textbf{EFT}};

\end{tikzpicture}
        }
    \caption{ \textit{Systematic computation of the connection matrix $\vb*{W}$}. 
    We first compute the NZ EFT basis perturbatively using the Born series~\cite{Caron-Huot:2025tlq}.
    The BHPT bases and the their connection matrix are computed using the MST method. By relating EFT and BHPT bases using transformation matrices, we can compute the EFT connection matrix through BHPT.
    }
    \label{fig:Interplay between EFT and BHPT}
\end{figure}
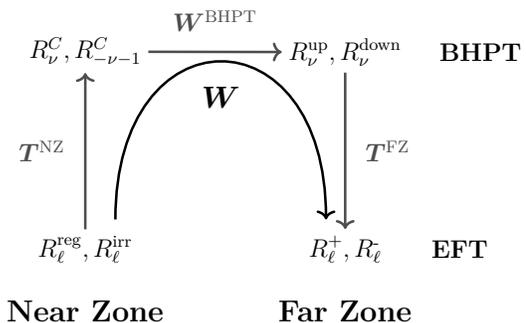

The application of BHPT to waveform is made possible by the following steps. 
First, we can set $d=4$ in the EFT NZ wavefunction after renormalization, such that the BHPT and EFT live in the same dimension.
Moreover, the radiative multipole moments and tidal interactions enter only through the NZ boundary condition at $r=0$. The homogeneous solutions are determined entirely by the Schwarzschild metric and are independent of any finite-size effects.
Therefore, the BHPT and EFT share the same wave equation for homogeneous solutions.

The BHPT solution bases can be obtained efficiently using the MST method as reviewed in the Supplemental Material. The solution bases are chosen differently between the BHPT and EFT.
A series of transformations are needed to derive the EFT connection matrix. We delineate our transformation procedure in Fig.~\ref{fig:Interplay between EFT and BHPT}.
In the NZ, we can derive the perturbative EFT solution basis $(\bar{R}^\tx{reg}_\ell, \bar{R}^\tx{irr}_\ell )$ from the Born series building from Eq.~\eqref{eq:wave_NZ} and $R_\ell = u_\ell/r$.
By comparing the EFT with the BHPT NZ solution basis  $(R^C_{\nu},R^C_{-\nu-1})$ with the expansion $GM\omega \ll \omega r \ll 1$ and matching the coefficients of each power of $r$,
we find the basis transformation
\begin{align}
    \begin{pmatrix}
        R^C_\nu \\ R^C_{-\nu-1}
    \end{pmatrix}
    =\vb*{T}^\tx{NZ}
    \begin{pmatrix}
        \bar{R}^\tx{reg}_\ell \\ \bar{R}^\tx{irr}_\ell
    \end{pmatrix},
    \label{eq:T_NZ}
\end{align}
where the NZ transition matrix $\vb*{T}^\tx{NZ}$ can only be a function of $GM\omega$ and $\mu$.
In the FZ, we simply need to align the BHPT basis $(R^\tx{up}_\nu,R^\tx{down}_\nu)$ with the asymptotic behaviors~\eqref{eq: EFT far zone asymptotic bc} of the EFT basis $(R^+_\ell, R^-_\ell)$ which leads to
\begin{align}
    \begin{pmatrix}
        R^+_\ell \\ R^-_\ell
    \end{pmatrix} = 
    \vb*{T}^\tx{FZ} 
    \begin{pmatrix}
        R^\tx{up}_\nu \\ R^\tx{down}_\nu
    \end{pmatrix},
    \label{eq:T_FZ}
\end{align}
The connection matrix can also be computed in the BHPT by the MST method,
\ie
\Big( \vb*{W}^\tx{BHPT} \Big)^T 
\begin{pmatrix}
R^\tx{up}_\nu \\ R^\tx{down}_\nu
\end{pmatrix}
=
\begin{pmatrix}
R^C_\nu \\R^C_{-\nu-1}
\end{pmatrix} \,.
\label{eq:W_BHPT}
\fe
Combining Eqs.~\eqref{eq:T_NZ}, \eqref{eq:T_FZ}, and \eqref{eq:W_BHPT}, we find the EFT connection matrix through
\ie
\vb*{W} = \left( \vb*{T}^\tx{FZ} \right)^{-1,T} \vb*{W}^\tx{BHPT} \left( \vb*{T}^\tx{NZ} \right)^{-1,T} \,.
\fe 
The matrix elements of $\vb*{W}$ then give the Sommerfeld factor~\eqref{eq:phi_final} and the Compton scattering $S$-matrix~\eqref{eq:S_matrix}.
Our method combines EFT with the MST method in BHPT and bypasses the need of nontrivial integration.

Using our method, we explicitly evaluate Sommerfeld factor and Compton $S$-matrix to $\order{G^{10}}$ for the partial wave $\ell =0,1,2$. Our Compton $S$-matrix agrees with the renormalized $\order{G^3}$ results in Eq.~(4.31) in~\cite{Correia:2025enx} and Eq.~(S14) in~\cite{Ivanov:2024sds}.
Our results to $\order{G^{5}}$ are given in the Supplemental Material and we leave the rest to the ancillary file \cite{Sommerfeld_data}.

\Section{Exact Renormalization Group of Multipole Moments} ---
The synergy between EFT and BHPT further illuminates the structures of the RG running.
Up to $\order{G^{2\ell+2}}$, there is no divergence associated with the tidal effect and the $\gamma_{11}$ in Eq.~\eqref{eq: RG of multipole moments} is fixed by the universal contribution from the Schwarzschild background.
As~\cite{Ivanov:2025ozg} has discovered, this universal contribution to $\gamma_{11}$ is given by the renormalized angular momentum $\nu$ in BHPT.

To cultivate the complete RG structure of $\bar{F}_\ell$ and $\bar{Q}_L$, we first notice that the RG equations~\eqref{eq: RG of multipole moments} can be solved in closed-forms in terms of $\vb*{\gamma}$, given in Eqs.~\eqref{eq: exact RG solution Q first} and \eqref{eq: exact RG solution F first}. 
In addition, we can connect $\vb*{\gamma}$ to the renormalized angular momentum beyond the universal part.
We start from the the RG equation of the EFT NZ basis,
$\left({d}/{d\log\mu}-\vb*{\gamma}^T\right) (\bar{R}^\tx{reg}_\ell, \bar{R}^\tx{irr}_\ell )^T =0$.
We notice that the NZ solutions in BHPT take the form
\begin{align}
    R^C_\nu &= e^{(\nu-\ell) \log(2\omega r)} ((2\omega r)^\ell + (2\omega r)^{-\ell-1} + \cdots)
     \\
    R^C_{-\nu-1} &= e^{-(\nu-\ell) \log(2\omega r)} (  (2\omega r)^\ell + (2\omega r)^{-\ell-1} + \cdots)\,, \nn
\end{align}
where we omit other integer powers of $\omega r$ in the parenthesis.
Since $(\bar{R}^\tx{reg}_\ell, \bar{R}^\tx{irr}_\ell )$ and $(R^C_{\nu},R^C_{-\nu-1})$ solve the same wave equation, their $\log r$ dependence should match after rotating the basis. 
But the logarithms appear as $\log(\mu r)$ in the EFT while as $\log(\omega r)$ in the BHPT. 
This implies that the NZ transition matrix takes the form 
$\vb*{T}^\tx{NZ} \rightarrow \tx{diag}( (\omega/\mu)^{\nu-\ell}, (\omega/\mu)^{-(\nu-\ell)})$ after a $\mu$-independent rotation. Since the BHPT basis is independent of $\mu$, we find
\ie
\tx{ the eigenvalues of}\; \vb*{\gamma} \;\tx{are}\; \pm( \ell - \nu ) .
\label{eq: anomalous gamma eigen value statement}
\fe
This statement holds even when the non-universal part is present.
We give the complete proof in the Supplemental Material.
Using the eigenvalues of $\vb*{\gamma}$, the complete solutions of the RG equations~\eqref{eq: RG of multipole moments} take the form
\begin{widetext}
\begin{align}
\bar{Q}_L(\omega,\mu)
=&
\bar{Q}_{L,0}(\omega)\,
\frac{\bar{F}_{\ell,-}(\omega)-\bar{F}_{\ell,+}(\omega)}{\bar{F}_{\ell,-}(\omega)-\bar{F}_{\ell,0}(\omega)+\bigl(\bar{F}_{\ell,0}(\omega)-\bar{F}_{\ell,+}(\omega)\bigr)\left( \tfrac{\mu}{\mu_0} \right)^{2(\nu(\omega)-\ell)}}
\left( \frac{\mu}{\mu_0} \right)^{(\nu(\omega)-\ell)} 
\label{eq:Q_RG}\\
\bar{F}_\ell(\omega,\mu)
=&
\bar{F}_{\ell,+}(\omega)
+ 
\frac{\bigl(\bar{F}_{\ell,0}(\omega)-\bar{F}_{\ell,+}(\omega)\bigr)\bigl(\bar{F}_{\ell,-}(\omega)-\bar{F}_{\ell,+}(\omega)\bigr)\left( \tfrac{\mu}{\mu_0} \right)^{2(\nu(\omega)-\ell)}}{\bar{F}_{\ell,-}(\omega)-\bar{F}_{\ell,0}(\omega)+\bigl(\bar{F}_{\ell,0}(\omega)-\bar{F}_{\ell,+}(\omega)\bigr)\left( \tfrac{\mu}{\mu_0} \right)^{2(\nu(\omega)-\ell)}},
\end{align}
\end{widetext}
where $\bar{Q}_{L,0}(\omega)$ and $\bar{F}_{\ell,0}(\omega)$ are the reference values of $\bar{Q}_L(\omega)$ and $\bar{F}_\ell(\omega)$ at the scale $\mu_0$. 
We have two fixed points $\bar{F}_{\ell,\mp}(\omega) \equiv
((\gamma_{22}-\gamma_{11}) \pm 2 (\ell-\nu))/(2\,c_\ell \gamma_{12})$. Eq.~\eqref{eq:Q_RG} explicitly shows the mixing between the radiative multipoles and the tidal response function which goes beyond the universal anomalous dimension of $\bar Q_L$ studied in \cite{Ivanov:2025ozg}. When $\mu/ \mu_0 \simeq \mathcal{O}(1)$, our new solution is consistent with \cite{Ivanov:2025ozg} as the tidal response dependence cancels out. But when there is a hierarchy $\mu/\mu_0 \gg \mathcal{O}(1)$, the contributions from tides become important.

\Section{Resummation Proposal} --- The exact solutions of the RG equations~\eqref{eq: RG of multipole moments} suggest an improved tail resummation in our scalar model. As noted in \cite{Ivanov:2025ozg}, the Sommerfeld factor $\mathcal{S}_{\ell}$ coincides with the tail resummation factor $T_{\ell}$ in the factorized MPM waveform. The RG running of the radiative multipoles efficiently resums the UV tail logs and captures the wave-amplitude correction due to scatterings against the curved background. Continuing from~\cite{Ivanov:2025ozg,Cipriani:2026xmx}, we further incorporate the mixing between the radiative multipoles and the tidal response, which leads us to following proposal
\begin{equation}
    |\mathcal{S}_\ell| = |\mathcal{S}_{\rm IR}| \times |\mathcal{S}_{\rm run}| \times |\mathcal{S}_{\rm rem}| ~,
\end{equation}
where each factor isolates a distinct physical mechanism. The infrared factor
\begin{equation}
    |\mathcal{S}|_{\rm IR} = \left|\frac{\Gamma(\nu(\omega)+1+2iG M \omega)\Gamma(2\ell+2)}{\Gamma(2\nu(\omega)+2)\Gamma(\ell+1)}\right| e^{\pi G M \omega} 
\end{equation}
captures the enhancement due to infrared loagirhtms. The running factor reads
\begin{widetext}
\begin{equation}
    |\mathcal{S}_{\rm run}| = \left|\frac{\bar{F}_{\ell,-}(\omega)-\bar{F}_{\ell,+}(\omega)}{\bar{F}_{\ell,-}(\omega)-\bar{F}_{\ell,0}(\omega)+\bigl(\bar{F}_{\ell,0}(\omega)-\bar{F}_{\ell,+}(\omega)\bigr) (\omega r_{\rm orb})^{2(\nu(\omega)-\ell)}}\right| (\omega r_{\rm orb})^{(\nu(\omega)-\ell)} ~,
\end{equation}
\end{widetext}
where $r_{\rm orb}$ is the binary orbital scale.
This factor improves \cite{Ivanov:2025ozg} in two ways: it promotes the universal anomalous dimension to the full BHPT value $\nu(\omega)-\ell$ to all orders in $G$, and it simultaneously resums the mixing with the tidal response. The remainder part $|\mathcal{S}_{\rm rem}|$ can be extracted order by order from the perturbative data of the Sommerfeld factor, which may be further improved using the Padé resummation. We record the remainder part for $\ell=1,2$ to $\mathcal{O}(G^6)$ in the Supplementary Material and leave the rest to the ancillary file~\cite{Sommerfeld_data}. 

Finally, we recall the exact relation~\eqref{eq:Sommerfeld_phase_Compton_phase} between the Sommerfeld phase and the elastic Compton phase shift. 
Whenever the absorptive sector is subleading, we can also use the elastic Compton phase shift, accessible directly from the scattering $S$-matrix~\cite{Caron-Huot:2025tlq,Bautista:2026qse,Bjerrum-Bohr:2026fhs,Ivanov:2026icp}, as a sharp proxy for the waveform phase in the resummation proposal.

\Section{Conclusion} ---
We have developed a systematic framework for computing the gravitational Sommerfeld factor of scalar perturbation from a generic compact source including tidal effects.
We derived a closed-form expression for the Sommerfeld factor in Eq.~\eqref{eq:phi_final}. 
We further streamlined the hybridization with BHPT, enabling the analytic computation of perturbative data to $\mathcal{O}(G^{10})$.
Along the way, we established the RG equation for radiative multipoles, including its mixing with tides, whose solution paves the way for improving future waveform models.

Our work opens several promising directions for future research. A key priority is the extension to spin-2 gravitational perturbations, which would enable a direct connection between the Sommerfeld factor and the gravitational waveform modes $h_{\ell m}$ used in data analysis. 
This generalization will require addressing important subtleties, including recoil effects on the worldline~\cite{Cheung:2023lnj,Kosmopoulos:2023bwc,Cheung:2024byb} and gauge dependence in spin-2 perturbations. 
Beyond this, realistic binary systems carry orbital angular momentum, leading to mixing between different quantum numbers $\ell$.
Capturing these effects naturally calls for a treatment in a Kerr background. 
Finally, a systematic investigation of different resummation proposals would be highly valuable.
Overall, our approach provides a natural starting point for developing a closer synergy between EFT and BHPT for precision gravitational-wave physics.

\Section{Acknowledgements} ---
We thank Maor Ben-Shahar, Miguel Correia, Giulia Isabella, Misha Ivanov, Mikhail Solon and Yue-Zhou Li for helpful discussion. Z.Z. would like to especially thank Matias Zaldarriaga for organizing the AI term at the IAS and for his support in facilitating the use of Claude Code.
CHC and CHS are supported by the Yushan Young Scholarship 112V1039 from
the Ministry of Education (MOE) of Taiwan, and also by the National Science and Technology Council (NSTC) grant 114L7329. 
CHS is also funded by the European Union
under ERC Starting Grant AmpEFT 101165689. 
Views and opinions expressed are however
those of the author(s) only and do not necessarily reflect those of the European Union or the European Research Council. Neither the European Union nor the granting authority can be
held responsible for them.


\bibliography{ref.bib}

\newpage 

\pagebreak
\widetext
\begin{center}
\textbf{\large Supplemental Material}
\end{center}
\setcounter{equation}{0}
\setcounter{figure}{0}
\setcounter{table}{0}
\setcounter{page}{1}
\makeatletter
\renewcommand{\theequation}{S\arabic{equation}}
\renewcommand{\thefigure}{S\arabic{figure}}

\section{Details on the Equation of Motion}
\label{app: EOM details}

In this appendix, we provide the explicit form of the wave equation and the potentials appearing therein. We work in $d = 4 - 2\epsilon$ dimensions with the Schwarzschild metric~\eqref{eq:Schwarzschild} and use the in-in formalism in the Schwinger--Keldysh basis. The complete action is $S = S_\tx{Bulk} + S_\tx{Love} + S_\tx{Source}$, where
\ie
S_\tx{Bulk} &=  - \int \dd[d]{x} \sqrt{-g}\, (\partial\phi_-)(\partial\phi_+)\,,\\
S_\tx{Love} &= \sum_{\ell,n} \frac{ \mu^{2\epsilon} }{\ell!} C_{\ell,n} \int_\gamma \dd{\tau}\, \nabla_L \phi_+ \nabla^L \partial^n_\tau\phi_-\,,\\
S_\tx{Source} &=  \sum_{\ell} \mu^{\epsilon} \int_\gamma \dd{\tau}\, Q^L(\tau) \nabla_L \phi_-\,.
\fe
The equation of motion for $\phi \equiv \phi_+$ follows from $\delta S / \delta \phi_- = 0$:
\ie
\bigg( \Box_\tx{Sch} + \mu^{2\epsilon} \sum_{n,\ell} \frac{(-1)^{\ell+n}}{\ell!} C_{\ell,n} \nabla_L \delta^{d-1}(\vb*{r}) \nabla^L \partial_t^n \bigg) \phi(x) = \mu^\epsilon \sum_{\ell} (-1)^{\ell+1} Q^L(t) \nabla_L \delta^{d-1}(\vb*{r})\,, \label{eq: box sch field equation}
\fe
where $\Box_\tx{Sch}$ is the scalar d'Alembertian in the Schwarzschild background,
\ie
\Box_\tx{Sch} = -\frac{1}{f}\partial_t^2 + \frac{1}{r^{d-2}} \partial_r \big( r^{d-2} f \,\partial_r \big) + \frac{\Delta_{S^{d-2}}}{r^2}\, ~,
\fe
where 
\begin{align}
    f(r) = 1 - \frac{2GM \mu^{2\epsilon}n_d}{r^{d-3}} \quad,\quad \tx{with} \quad n_d = \frac{4\pi^{\frac{3-d}{2}} \Gamma(\frac{d-1}{2})}{d-2} ~,
\end{align}
and $\Delta_{S^{d-2}}$ is the Laplacian on the unit $(d-2)$-sphere. The source term on the right-hand side encodes the multipole moments of the compact object, while the tidal potential on the left-hand side captures the response of the compact object to external perturbations.

\paragraph{Radial equation.}
To derive Eq.~(\ref{eq: radial equation with source}) in the main text, we first decompose $\phi$ into spherical harmonics:
\begin{equation}
    \phi(r,\Omega) = e^{-i \omega t} \sum_{\ell,m} R_{\ell m}(r) Y_{\ell m}(\Omega) \equiv e^{-i \omega t} \sum_{\ell, m} \frac{u_{\ell m}(r)}{r^{(d-2)/2}} Y_{\ell m}(\Omega)\,.
\end{equation}
Then, the equation of motion reduces to a radial equation for $u_{\ell m}(r)$:
\begin{equation}
    \begin{aligned}
    & \quad r^{-1+\epsilon} f(r) u_{\ell m}''(r) + r^{-1+\epsilon} f'(r) u_{\ell m}'(r) + r^{-1+\epsilon}\frac{\omega^2}{f(r)} u_{\ell m}(r) + r^{\epsilon -3} \Big(r (\epsilon -1) f'(r)-(\epsilon -1) \epsilon  f(r)-\ell (\ell -2 \epsilon +1)\Big) u_{\ell m}(r) \\
    & = V_{\rm Love}(r) R_{\ell m}(r) + \mu^{\epsilon} \sum_{\ell} (-1)^{\ell+1} Q^L(\omega) \int_\Omega Y_{\ell m}^* \nabla_L \delta^{d-1}(\vb*{r})\,,
    \end{aligned}
\end{equation}
where the tidal potential $V_{\rm Love}(r)$ is given by
\begin{equation}
    V_{\rm Love}(r) = - \sum_{\ell,n} \frac{(-1)^{\ell}}{\ell!} C_{\ell,n} \mu^{2\epsilon} \nabla_L \delta^{d-1}(\vb*{r}) \nabla^L (i \omega)^{n} \equiv -\mu^{2 \epsilon} \sum_{\ell} F_{\ell}(\omega) \frac{(-1)^{\ell}}{\ell!} \nabla_{L} \delta^{d-1}(\mathbf{r}) \nabla_{L} ~,
\end{equation}
with the tidal response function $F_\ell(\omega) = \sum_n C_{\ell,n} (i\omega)^n$. To clean up second-derivative term in the equation, we divide both sides by $r^{-1+\epsilon} f(r)$ and find
\begin{equation}
    \begin{aligned}
    & \quad u_{\ell m}''(r) + \frac{f'(r)}{f(r)} u_{\ell m}'(r) + \frac{\omega^2}{f(r)^2} u_{\ell m}(r) + \frac{1}{r^2 f(r)} \Big(r (\epsilon -1) f'(r)-(\epsilon -1) \epsilon  f(r)-\ell (\ell -2 \epsilon +1)\Big) u_{\ell m}(r) \\
    & = r^{1-\epsilon} \frac{V_{\rm Love}(r)}{f(r)} R_{\ell m}(r) + r^{1-\epsilon} \frac{\mu^{\epsilon}}{f(r)} \sum_{\ell} (-1)^{\ell+1}  Q^L(\omega) \int_\Omega Y_{\ell m}^* \nabla_L \delta^{d-1}(\vb*{r})\,.
    \end{aligned}
\end{equation}
Since the tidal response function and the source term are localized at the origin, we can further simplify the equation by replacing $f(r)$ with $1$ on the right-hand side. This leads to the final form of the radial equation:
\begin{equation}
    \begin{aligned}
    \left[\frac{d^2}{d r^2}-\frac{(\ell-\epsilon)(\ell-\epsilon+1)}{r^2}+\omega^2\right] u_{\ell m}(r) = V_{\rm Grav}(r) u_{\ell m}(r) + r^{1-\epsilon} \Big( V_{\rm Love}(r) R_{\ell m}(r) + \mu^{\epsilon} \sum_{\ell} (-1)^{\ell+1}  Q^L(\omega) \int_\Omega Y_{\ell m}^* \nabla_L \delta^{d-1}(\vb*{r}) \Big)\,.
    \end{aligned}
\end{equation}
with the effective gravitational potential defined as~\cite{Caron-Huot:2025tlq}
\begin{equation}\begin{aligned}
& V_{\text {Grav }}(r)=\sum_{n=1}^{\infty}\left(\frac{2 G M n_d \mu^{2 \epsilon}}{r^{1-2 \epsilon}}\right)^n\left[\frac{2 \epsilon-1}{r} \frac{d}{d r}+\frac{\ell^2+\ell+1-\epsilon(3+2 \ell)+2 \epsilon^2}{r^2}-(n+1) \omega^2\right] ~.
\end{aligned}
\end{equation}
Once transformed into the STF basis, we recover Eq.~\eqref{eq: radial equation with source}.

\section{Spherical Harmonic Basis and STF Basis}\label{app: Spherical Harmonic and STF basis review}

In the previous appendix, the radial equation was written in the spherical harmonic basis. However, the localized tidal Love term and the multipole source are more naturally expressed in terms of STF tensors. In this appendix, we therefore collect the identities that convert between the spherical harmonic basis and the STF basis, and then show the derivative formulas needed for deriving the near-zone boundary condition.

We begin by defining the scalar spherical harmonics $Y^{(d)}_{\ell \vb*{m}}(\hat{\vb*{r}})$ on the unit sphere $S^{d-2}$ as an orthonormal basis of eigenfunctions of the spherical Laplacian,
\begin{align}
\Delta_{S^{d-2}} Y^{(d)}_{\ell \vb*{m}}(\hat{\vb*{r}})
=
-\ell(\ell+d-3)\,Y^{(d)}_{\ell \vb*{m}}(\hat{\vb*{r}}),
\end{align}
\begin{align}
\int \dd{\Omega_{d-2}}\,
Y^{*(d)}_{\ell \vb*{m}}(\hat{\vb*{r}})
Y^{(d)}_{\ell' \vb*{m}'}(\hat{\vb*{r}})
=
\delta_{\ell\ell'}\delta_{\vb*{m}\vb*{m}'}.
\end{align}
For fixed $\ell$, the label $\vb*{m}$ runs over a basis of the degenerate eigenspace. A convenient choice is the basis adapted to the multiplicity-free branching chain
\begin{align}
SO(d-1)\supset SO(d-2)\supset \cdots \supset SO(3)\supset SO(2),
\end{align}
with 
\begin{equation}
    {\rm Res}_{SO(N)}^{SO(N-1)} V_{\ell_n} = \bigoplus_{\ell_{n-1}=0}^{\ell_n} V_{\ell_{n-1}} ~,
\end{equation}
where $V_{\ell_n}$ is the irreducible representation for $SO(N)$ and $V_{\ell_{n-1}}$ is the irreducible representation for $SO(N-1)$. Therefore, it is convenient to introduce a set of quantum numbers
\begin{align}
\vb*{m}\equiv(\ell_{d-3},\ell_{d-4},\dots,\ell_2,m)
\end{align}
to efficiently capture the multiplicity-free branching chain structure with
\begin{align}
\ell \ge \ell_{d-3}\ge \ell_{d-4}\ge \cdots \ge \ell_2 \ge |m|,
\end{align}
where $\ell_j\in \mathbb Z_{\ge 0}$ for $j=2,\dots,d-3$ and $m\in \mathbb Z$. In $d=4$, this reduces to the familiar $SO(3)\supset SO(2)$ labeling by a single magnetic quantum number $m=-\ell,-\ell+1,\dots,\ell$. We only need one representation-theoretic fact here: this eigenspace furnishes the rank-$\ell$ symmetric trace-free irrep of $SO(d-1)$. Equivalently, it is in one-to-one correspondence with rank-$\ell$ STF tensors $T_{i_1\cdots i_\ell}$, which has the properties,
\begin{align}
T_{i_1\cdots i_\ell} = T_{(i_1\cdots i_\ell)},
\qquad
\delta^{i_1 i_2} T_{i_1 i_2 i_3\cdots i_\ell} = 0.
\end{align}
For convenience, we denote indices collectively by the notation $L \equiv i_1\cdots i_\ell$. The degeneracy of scalar harmonics with angular momentum $\ell$ is therefore the dimension of this irreducible representation,
\begin{align}
\Tr 1_\ell
=
\frac{(d+2\ell-3)\Gamma(d+\ell-3)}{\Gamma(d-2)\Gamma(\ell+1)}\,,
\end{align}
where $1_\ell$ denotes the identity operator on the $\ell$-th irreducible representation.

Thus the spherical harmonic basis and the STF basis are simply two different choices of basis for the same irreducible representation, with $\hat{\vb*{r}}_L$ denoting the STF product of $\ell$ unit vectors.
The scalar spherical harmonics form a complete orthonormal basis on $S^{d-2}$, so
\begin{align}
\sum_{\ell=0}^{\infty}\sum_{\vb*{m}}
Y^{*(d)}_{\ell \vb*{m}}(\hat{\vb*{r}})\,
Y^{(d)}_{\ell \vb*{m}}(\hat{\vb*{r}}')
=
\delta^{(d-2)}(\hat{\vb*{r}},\hat{\vb*{r}}'),
\end{align}
where $\delta^{(d-2)}(\hat{\vb*{r}},\hat{\vb*{r}}')$ is the delta function on the unit sphere, normalized by
\begin{align}
\int \dd{\Omega'_{d-2}}\,\delta^{(d-2)}(\hat{\vb*{r}},\hat{\vb*{r}}')\,f(\hat{\vb*{r}}')
=
f(\hat{\vb*{r}}).
\end{align}
If instead one keeps $\ell$ fixed, then the sum over $\vb*{m}$ gives the projector onto the $\ell$-th harmonic subspace,
\begin{align}
K_\ell(\hat{\vb*{r}},\hat{\vb*{r}}')
\equiv
\sum_{\vb*{m}}
Y^{*(d)}_{\ell \vb*{m}}(\hat{\vb*{r}})\,
Y^{(d)}_{\ell \vb*{m}}(\hat{\vb*{r}}'),
\end{align}
which is a smooth rotationally invariant function of $\hat{\vb*{r}}\cdot \hat{\vb*{r}}'$ rather than a delta function. To derive its explicit form, note that for fixed $\hat{\vb*{r}}'$ the function $K_\ell(\hat{\vb*{r}},\hat{\vb*{r}}')$ lies in the $\ell$-th harmonic subspace as a function of $\hat{\vb*{r}}$. Moreover,
\begin{align}
K_\ell(R\hat{\vb*{r}},R\hat{\vb*{r}}')
=
K_\ell(\hat{\vb*{r}},\hat{\vb*{r}}')
\qquad \forall R\in SO(d-1),
\end{align}
so it can only depend on the invariant
\begin{align}
y \equiv\hat{\vb*{r}}\cdot \hat{\vb*{r}}'.
\end{align}
Defining $K_\ell(y) \equiv K_\ell(\hat{\vb*{r}},\hat{\vb*{r}}')$ and using
\begin{align}
\Delta_{S^{d-2}}K_\ell(y)
=
(1-y^2)K_\ell''(y)-(d-1)x K_\ell'(y),
\end{align}
the harmonic eigenvalue equation becomes
\begin{align}
(1-x^2)K_\ell''(x)-(d-1)x K_\ell'(x)+\ell(\ell+d-3) K_\ell(x)=0.
\end{align}
This is the Gegenbauer differential equation with $\lambda=(d-3)/2$, so regularity on the sphere implies
\begin{align}
K_\ell(\hat{\vb*{r}},\hat{\vb*{r}}')
=
A_\ell\,C_\ell^{\frac{d-3}{2}}(x).
\end{align}
The coefficient $A_\ell$ is fixed by setting $\hat{\vb*{r}}'=\hat{\vb*{r}}$ and integrating over the sphere:
\begin{align}
S_{d-2}\,A_\ell\,C_\ell^{\frac{d-3}{2}}(1)
=
\sum_{\vb*{m}}\int \dd{\Omega_{d-2}}\,\big|Y^{(d)}_{\ell \vb*{m}}(\hat{\vb*{r}})\big|^2
=
\Tr 1_\ell.
\end{align}
Using $C_\ell^\lambda(1)=\Gamma(\ell+2\lambda)/[\Gamma(2\lambda)\Gamma(\ell+1)]$ together with the expression for $\Tr 1_\ell$, one finds
\begin{align}
A_\ell=\frac{2\ell+d-3}{(d-3)S_{d-2}}.
\end{align}
Therefore the addition theorem reads
\begin{align}
K_\ell(\hat{\vb*{r}},\hat{\vb*{r}}')
=
\frac{2\ell+d-3}{(d-3)S_{d-2}}\,
C_\ell^{\frac{d-3}{2}}\!\left(\hat{\vb*{r}}\cdot \hat{\vb*{r}}'\right),
\end{align}
where $C_\ell^\lambda(x)$ is the Gegenbauer polynomial. Setting $\hat{\vb*{r}}'=\hat{\vb*{r}}$ and using
\begin{align}
C_\ell^\lambda(1)=\frac{\Gamma(\ell+2\lambda)}{\Gamma(2\lambda)\Gamma(\ell+1)},
\end{align}
we obtain
\begin{align}
\sum_{\vb*{m}} Y^{*(d)}_{\ell \vb*{m}}(\hat{\vb*{r}})\,Y^{(d)}_{\ell \vb*{m}}(\hat{\vb*{r}})
=
\frac{\Tr 1_\ell}{S_{d-2}}. \label{eq: sum of two spherical harmonics with the same position}
\end{align}

\paragraph{Spherical Harmonics and STF Tensors.}
We now review the explicit map between the spherical harmonic in the $(\ell, \vb*{m})$ basis and the STF basis. The spherical harmonics in the STF basis used in the main text are simply
\begin{align}
    Y_L \equiv \hat{r}_L \,.
\end{align}
We start from the decomposition of the spherical harmonics,
\begin{align}
    Y^{(d)}_{\ell \vb*{m}} = \mathcal{Y}^L_{\ell \vb*{m}} \hat{r}_L  \label{eq: spherical harmonic project on STF}
\end{align}
where $\mathcal Y^L_{\ell \vb* m}$ is a constant rank-$\ell$ STF tensor. Here $\delta^{LL'}_{\tx{STF}}$ denotes the identity operator on the rank-$\ell$ STF subspace,
\begin{align}
\delta^{LL'}_{\tx{STF}}\, T_{L'} = T^L,
\qquad
\delta^{LL'}_{\tx{STF}}
=
\delta^{\langle i_1}_{\langle j_1}\cdots \delta^{\,i_\ell\rangle}_{j_\ell\rangle},
\end{align}
where the brackets denote STF projection on the $i$- and $j$-indices separately. Writing $\lambda=(d-3)/2$, one may also express this projector explicitly as
\begin{align}
\delta_{\tx{STF}}^{i_1\cdots i_\ell,\; j_1\cdots j_\ell}
=
\sum_{k=0}^{\lfloor \ell/2\rfloor}
\frac{(-1)^k\,\ell!\,\Gamma(\ell-k+\lambda)}
{4^k\,k!\,(\ell-2k)!\,\Gamma(\ell+\lambda)}
\,
\mathcal S_i \mathcal S_j
\Big[
\delta^{i_1 i_2}\cdots \delta^{i_{2k-1}i_{2k}}
\delta^{j_1 j_2}\cdots \delta^{j_{2k-1}j_{2k}}
\delta^{i_{2k+1}}_{j_{2k+1}}\cdots \delta^{i_\ell}_{j_\ell}
\Big],
\end{align}
where $\mathcal S_i$ and $\mathcal S_j$ denote symmetrization with unit weight over the $i$- and $j$-indices separately, with all terms absent when their upper index is smaller than the lower one. For example, for $\ell=2$ this reduces to
\begin{align}
\delta_{\tx{STF}}^{ij,kl}
=
\frac12\left(\delta^{ik}\delta^{jl}+\delta^{il}\delta^{jk}\right)
-\frac{1}{d-1}\delta^{ij}\delta^{kl}.
\end{align}
The bilinear form
\begin{align}
\int \dd{\Omega_{d-2}}\,\hat r_L \hat r_{L'}
\end{align}
is $SO(d-1)$ invariant on the irreducible rank-$\ell$ STF representation, so Schur's lemma implies
\begin{align}
\int \dd{\Omega_{d-2}}\,\hat r_L \hat r_{L'}
=
A_\ell\,\delta^{LL'}_{\tx{STF}}
\end{align}
for some constant $A_\ell$. Contracting with $\delta^{LL'}_{\tx{STF}}$ gives
\begin{align}
A_\ell\, \Tr 1_\ell
=
\delta^{LL'}_{\tx{STF}}\int \dd{\Omega_{d-2}}\,\hat r_L \hat r_{L'}
=
\int \dd{\Omega_{d-2}}\, \hat r^L \hat r_L
=
S_{d-2}\,\hat r^L \hat r_L .
\end{align}
Using
\begin{align}
\hat r^L \hat r_L
=
\frac{\sqrt{\pi}\,\Gamma(d+\ell-3)}
{2^{d+\ell-4}\Gamma\!\left(\frac d2-1\right)\Gamma\!\left(\frac{d-3}{2}+\ell\right)},
\end{align}
together with the expression for $\Tr 1_\ell$, one finds
\begin{align}
A_\ell=\frac{\ell!}{c_\ell}
\qquad \tx{with} \qquad
c_\ell = \frac{2^\ell}{2\pi^{\frac{d-1}{2}}}\Gamma\!\left(\frac{d-1}{2}+\ell\right).
\end{align}
Therefore the STF basis obeys
\begin{align}
\int \dd{\Omega_{d-2}}\,\hat r_L \hat r_{L'}
=
\frac{\ell!}{c_\ell}\,\delta^{LL'}_{\tx{STF}} .
\end{align}
Multiplying \eqref{eq: spherical harmonic project on STF} by $Y^{*(d)}_{\ell \vb* m}$, integrating over the sphere, and using the previous identity gives
\begin{align}
\delta_{\vb* m \vb* m'}
=
\int \dd{\Omega_{d-2}}\,
Y^{*(d)}_{\ell \vb* m} Y^{(d)}_{\ell \vb* m'}
=
\frac{\ell!}{c_\ell}\,\mathcal Y^{*L}_{\ell \vb* m}\mathcal Y^{L}_{\ell \vb* m'} ,
\end{align}
which is the orthonormality relation for the coefficients $\mathcal Y^L_{\ell \vb* m}$. Since $\mathcal Y^L_{\ell \vb* m}$ is the change-of-basis matrix between two bases of the same irreducible representation, the inverse relation is
\begin{align}
\sum_{\vb* m}\mathcal Y^{*L}_{\ell \vb* m}\mathcal Y^{L'}_{\ell \vb* m}
=
\frac{c_\ell}{\ell!}\,\delta^{LL'}_{\tx{STF}} . \label{eq: sum of mathcal Y}
\end{align}
Using this in \eqref{eq: spherical harmonic project on STF}, we obtain the inverse map
\begin{align}
\hat r_L = \frac{\ell!}{c_\ell} \sum_{\vb* m} \mathcal Y^{*L}_{\ell \vb* m} Y^{(d)}_{\ell \vb* m} \label{eq: STF basis in term of Spherical Harmonic basis}
\end{align}

\paragraph{Basis change for components.}
With these relations, any function may be expanded in either basis,
\begin{align}
    f(\vb*{r}) = \sum_{\ell = 0}^\infty \sum_{\vb*{m}} f_{\ell \vb*{m}}(r) Y^{(d)}_{\ell \vb*{m}} = \sum_{\ell = 0}^\infty f_L(r) \hat{\vb*{r}}^L .
\end{align}
Using the orthogonality relations, we have
\begin{align}
    f_{\ell \vb*{m}} (r) = \int \dd{\Omega_{d-2}} Y^{(d)*}_{\ell \vb*{m}} f(\vb*{r})
    \quad,\quad
    f_L (r) = \frac{c_\ell}{\ell!} \int \dd{\Omega_{d-2}} \hat{\vb*{r}}_L f(\vb*{r}) .
\end{align}
The transition relation between the two bases is therefore
\begin{align}
    f_L(r) = \sum_{\vb{m}} f_{\ell \vb*{m}} (r) \mathcal{Y}^L_{\ell \vb*{m}}
    \quad,\quad
    f_{\ell \vb*{m}} (r) = \frac{\ell!}{c_\ell} f_L (r) \mathcal{Y}^{*L}_{\ell \vb*{m}} . \label{eq: STF and spherical harmonics basis component transition}
\end{align}

\paragraph{Derivative identities.}
We now show the derivative identities that are used in the derivation of the localized tidal Love and source terms in the earlier sections. We first derive
\begin{align}
    \partial_L \frac{1}{r^{d-3}} = (-2)^\ell \left( \frac{d-3}{2} \right)_\ell \frac{\hat{\vb*{r}}_L}{r^{d+\ell-3}} \label{eq: STF derivative on 1/r}
\end{align}
where $(z)_n = \Gamma(z+n)/\Gamma(z)$ is the Pochhammer symbol. This formula is useful for the derivation of the boundary condition \eqref{eq: near zone boundary condition}. To prove it, we use \cite{Blanchet:1985sp}
\begin{align}
    \partial_L f(r) = \hat{\vb*{r}}_L r^\ell \left( \frac{1}{r} \dv{}{r} \right)^\ell f(r) .
\end{align}
Setting $f(r) = r^{3-d}$ yields \eqref{eq: STF derivative on 1/r}. Moreover,
\begin{align}
    \nabla^2 \frac{1}{r^{d-3}} = -(d-3)S_{d-2} \delta^{d-1}(\vb*{r}) = -\frac{4\pi^{ \frac{d-1}{2} } }{\Gamma \left( \frac{d-3}{2} \right) } \delta^{d-1}(\vb*{r}) .
\end{align}
Combining this with \eqref{eq: STF derivative on 1/r}, we conclude
\begin{align}
    \nabla^2 \frac{ \hat{\vb*{r}}_L }{ r^{d+\ell-3} } = - \frac{1}{ (-2)^\ell \left( \frac{d-3}{2} \right)_\ell } \frac{4\pi^{ \frac{d-1}{2} } }{\Gamma \left( \frac{d-3}{2} \right) } \partial_L \delta^{d-1}(\vb*{r}) \label{eq: nabla^2 on 1/r^d+l-3}
\end{align}
which is the identity used in the derivation of the near-zone boundary condition.

\section{Determinant of the Connection Matrix}\label{app: Determinant of the Connection Matrix}

In this appendix, we are going to prove that the determinant of the connection matrix defined in (\ref{eq: EFT connection matrix def}) is:
\begin{align}
    \det \vb*{W} = \frac{1}{2i\omega}
    \label{eq:W_det}
\end{align}
The connection matrix can be written as the usual Wronskian, defined as
\begin{align}
    W[f,g](r) = f(r) g'(r) - f'(r) g(r) \,.
\end{align}
Suppose both $f,g$ in the Wronskian $W[f,g]$ satisfy the differential equation, $y'' + q(r) y' + p(r) y = 0$, then the Wronskian as a function of $r$ satisfies the Abel's identity
\begin{align}
    W'[f,g](r) = - p(r) W[f,g](r) \,.
    \label{eq:Abel_id}
\end{align}
Through out the paper, we use bold font for the connection matrix and its matrix element as $W_{ij}$ to distinguish them from the Wronskian. We also drop the argument $r$ of the Wronkskian unless it is fixed to a particular value.
Using this definition and Eq.~\eqref{eq: EFT connection matrix def}, we find
\begin{align}
    W_{11} = \frac{W[u^-,u^\tx{reg}]}{W[u^-,u^+]} \quad,\quad W_{12} = \frac{W[u^-,u^\tx{irr}]}{W[u^-,u^+]} \quad,\quad
    W_{21} = \frac{W[u^+,u^\tx{reg}]}{W[u^+,u^-]} \quad,\quad W_{22} = \frac{W[u^+,u^\tx{irr}]}{W[u^+,u^-]}
\end{align}
By direct calculation, we can see that
\begin{align}
    \det \vb*{W} = \frac{W[u^\tx{irr},u^\tx{reg}]}{W[u^-,u^+]}\,.
    \label{eq:detW_Wronskian}
\end{align}
Let us first show that the ratio of two Wronskians is a constant in $r$. Considering two Wronskians, $W_a$ and $W_b$, whose arguments satisfying the same ODE, $y'' + q(r) y' + p(r) y = 0$, we can check that 
\begin{align}
    \dv{}{r} \frac{W_a}{W_b} = \frac{W'_a W_b - W_a W'_b}{W_b^2} = 0 
\end{align}
by Abel's identity. Since the ratio is a constant, we can evaluate it at $r = \infty$ for the sake of convenience. From the (\ref{eq: EFT far zone asymptotic bc}), we find $W[u^-, u^+](\infty) = 2i\omega$. In addition, we also have
\begin{align}
    W[u^\tx{irr},u^\tx{reg}](\infty) =& W[u^\tx{irr},u^\tx{reg}](0) \times \exp( \int_0^\infty \dd{r'}\sum_{n=1}^\infty \left( \frac{2GM\mu^{2\epsilon}(2\epsilon-1)}{r'^{1-2\epsilon}} \right)^n ), \nn \\
    =& W[u^\tx{irr},u^\tx{reg}](0) \,,
    \label{eq:Wronskian_constant}
\end{align}
where we use that the exponential is unity under dimensional regularization. We can then evaluate the Wronskian from the near zone
\begin{equation}
\begin{split}
    W[u^\tx{irr},u^\tx{reg}](\infty)
    = &W[u^\tx{irr},u^\tx{reg}](0) \\
    = & W \left[ \frac{\mu^{\epsilon}}{2\ell-2\epsilon+1} r^{-\ell + \epsilon}, \mu^{-\epsilon} r^{\ell+1-\epsilon} \right]\bigg\vert_{r = 0}  + \tx{gravitational corrections} \\
    =& 1 + \tx{gravitational corrections} 
\end{split}
\end{equation}
However, by inspecting the form of $\Vgrav$, we find that the gravitational correction always contains $r^{b\epsilon}$, with some positive integers $b$, which is set to $0$ at $r = 0$ by dimensional regularization. 
Thus, we find $W[u^\tx{irr},u^\tx{reg}](\infty)=1$ and $W[u^-, u^+](\infty) = 2i\omega$. Using Eq.~\eqref{eq:detW_Wronskian}, we complete the proof of Eq.~\eqref{eq:W_det}.

\section{Derivation of the Sommerfeld Factor from Boundary Conditions}\label{app: solving the wave form}

In this appendix, we complement the detailed derivation of the waveform and the Sommerfeld factor from the boundary conditions, including the proof of Eq.~(\ref{eq: near zone boundary condition}) and the Green's function method.

\paragraph{Boundary Condition Method.} 
We first derive the boundary condition (\ref{eq: near zone boundary condition}). Afterward, we show how to solve for the waveform $\Aout$ explicitly. As mentioned in the main text, the boundary condition originates from the presence of the Dirac delta function. Consequently, it suffices to focus on this localized contribution. Moreover, gravitational effects enter universally in the form $1/r^{1-2\epsilon}$, which vanishes as we set $r = 0$ due to dimensional regularization. Therefore, in deriving the boundary condition, we are able to ignore gravitational potential and consider the following equation instead
\begin{align}
    \left( \nabla_\tx{Flat}^2 +  \mu^{2\epsilon} \sum_{\ell = 0}^\infty \frac{(-1)^\ell}{\ell!} F_\ell (\omega) \partial_L \delta^{d-1} (\vb*{r}) \partial_L \right) \phi(\vb*{r}) =  \mu^{\epsilon} \sum_{\ell = 0}^\infty (-1)^{\ell+1} \QL^L \partial_L \delta^{d-1} (\vb*{r}) ,
\end{align}
which is obtained from (\ref{eq: box sch field equation}) by setting $G = 0$ and performing a Fourier transform. We also drop the $\omega^2$ term since it doesn't affect the boundary condition. We start with the first term and show that the Laplacian acting on $\uirr$ produces a Dirac delta function. As explained above, since gravitational effects do not influence the boundary condition, we can also set $G = 0$ in the wave function. Let $u^\tx{0,reg}_\ell = \ureg |_{G=0}$ and $u^\tx{0,irr}_\ell = \uirr |_{G=0}$. Then, we have
\begin{equation}
\begin{split}
    \nabla_\tx{Flat}^2 \phi(\vb*{r}) &= \nabla_\tx{Flat}^2 \sum_{\ell = 0}^\infty  \left( B_L^\tx{reg} u^\tx{0,reg}_\ell (r) + B_L^\tx{irr} u^\tx{0,irr}_\ell (r) \right) \frac{ \hat{\vb*{r}}_L }{r^{(d-2)/2}} \\
    &= B_L^\tx{irr} \sum_{\ell = 0}^\infty  \nabla_\tx{Flat}^2 \left( \frac{\mu^{\epsilon}}{2\ell - 2\epsilon + 1} \frac{\hat{\vb*{r}}_L}{r^{\ell+1-2\epsilon}} \right) =  \sum_{\ell = 0}^\infty \frac{(-1)^{\ell+1}}{c_\ell} \mu^{\epsilon} B^\tx{irr}_L \partial_L \delta^{d-1} (\vb*{r})\,,
\end{split}
\end{equation}
where we have used Eq.~\eqref{eq: nabla^2 on 1/r^d+l-3}. In the second equality, we drop $u^\tx{0,reg}_\ell$ since it satisfies the Laplacian equation and does not contain any singularity. Next, we compute the Dirac delta function produced by the tidal potential
\begin{equation}
\begin{split}
     & \mu^{2\epsilon} \sum_{\ell = 0}^\infty \frac{(-1)^\ell}{\ell!} F_\ell (\omega) \partial_L \delta^{d-1} (\vb*{r}) \partial_L  \phi(\vb*{r}) \\
     = & \mu^{2\epsilon} \sum_{\ell = 0}^\infty \frac{(-1)^\ell}{\ell!} F_\ell (\omega) \partial_L \delta^{d-1} (\vb*{r}) \partial_L \sum_{\ell' = 0}^\infty \left( B_{L'}^\tx{reg} u^\tx{0,reg}_{\ell'}(r) + B_{L'}^\tx{irr} u^\tx{0,irr}_{\ell'} (r) \right) \frac{ \hat{\vb*{r}}_{L'} }{r^{(d-2)/2}} \\
     =&  \sum_{\ell = 0}^\infty (-1)^\ell F_\ell(\omega) \mu^{\epsilon}  B^\tx{reg}_L  \partial_L \delta^{d-1}(\vb*{r}) \,,
\end{split}
\end{equation}
where $u^\tx{0,irr}_\ell$ doesn't contribute since it is evaluated at $r = 0$ due to Dirac delta function and vanishes thanks to dimensional regularization. Finally, combining first, second and the sources terms and requiring $\partial_L \delta^{d-1}(\vb*{r})$ to be canceled, we then finish the proof on the boundary condition~\eqref{eq: near zone boundary condition}.

We can solve $\Breg$ and $\Birr$ from the near-zone and far-zone boundary conditions
\begin{equation}
\begin{cases}
    &c_\ell^{-1} \Birr - F_\ell(\omega) \Breg = Q_L \\[3pt]
    &W_{21} \Breg + W_{22} \Birr = A^-_L = 0\,.
\end{cases}
\end{equation}
Furthermore, since $\Aout$ can be related to $\Breg$ and $\Birr$ through the connection matrix, this leads to
\ie
    \Aout = -  \det \vb*{W} \frac{ c_\ell Q_L }{ W_{21} + c_\ell F_\ell(\omega) W_{22} } = 
    \frac{i}{2\omega}\,\frac{ c_\ell Q_L }{ W_{21} + c_\ell F_\ell(\omega) W_{22} }
\fe
where we have used $\det \vb*{W} = 1/(2i\omega)$. Combining the Eqs.\eqref{eq: scalar wave spherical decomposition}, \eqref{eq: EFT far zone asymptotic bc}, and \eqref{eq:wave_FZ}, we complete the derivation of Eq.~(\ref{eq:phi_final}).

\paragraph{Green's function method.}
We first derive the Green’s function and the associated boundary conditions, and then calculate the waveform. The Green's function is defined by the following equation
\ie
    \left( \dv[2]{}{r} + \omega^2 - \frac{ (\ell-\epsilon)(\ell-\epsilon+1) }{ r^2 } - V(r) \right) G_\ell (r,r') = \delta(r-r') \,.
\fe
We can construct the Green's function from two homogeneous solutions $u^>_\ell$ and $u^<_\ell$ of the differential equations as
\ie
    G_\ell (r,r') = \frac{ u^<_\ell (r_<) u^>_\ell (r_>) }{ W[u^<_\ell, u^>_\ell](r') } \,,
    \label{eq:Green_fx}
\fe 
where we define $r_> = \max \{ r,r'\}$ and $r_< = \min \{ r,r'\}$.
We can impose the boundary conditions on $G_\ell$ by demanding the correct boundary behaviors of $u^>_\ell$ and $u^<_\ell$.
Since we only consider the emission process, $u^>_\ell$ should only be outgoing for large $r$. Consequently, we should pick $u^>_\ell = \uout$ in Eq.~\eqref{eq:wave_FZ}.
On the other hand, $u^<_\ell$ should obey the boundary condition that encodes the correct boundary behavior at $r=0$. 
Since $u^<_\ell$ satisfies the homogeneous differential equation, we should use the source-free near-zone boundary condition $B^\text{irr}_\ell/B^\text{reg}_\ell = c_\ell F_\ell$ from Eq.~\eqref{eq: near zone boundary condition} when we write $u^<_\ell$ in the near-zone basis, $u^<_\ell = B^\text{reg}_\ell \ureg + B^\text{irr}_\ell \uirr$.
For convenience, we can choose to normalize $u^<_\ell$ such that $u^<_\ell = A^+_\ell \uout + \uin$ when it is written in the far-zone basis.
Using the connection matrix and the source-free near-zone boundary condition, we can fix $u^<_\ell$ entirely. In summary, our Green's function in~Eq.~\eqref{eq:Green_fx} is built from
\ie
    u^>_\ell(r) &= \uout(r) \\
    u^<_\ell(r) &= \left(\frac{W_{11}+c_\ell F_\ell W_{12}}{W_{21}+c_\ell F_\ell W_{22}} \right) \uout(r) + \uin(r) = \frac{1}{W_{21}+c_\ell F_\ell W_{22}} \left(\ureg(r)+c_\ell F_\ell \uirr(r) \right)\,.
    \label{eq:Green_fx_u}
\fe
To calculate the Wronskian $W[u^<_\ell, u^>_\ell](r')$, we can use the above decomposition and the Abel's identity and $\Vgrav$ to show that
\begin{align}
    W[u^<_\ell, u^>_\ell] (r') 
    &= W[u^-_\ell, u^+_\ell] (r') \nn \\
    &= W[u^-_\ell, u^+_\ell] (\infty) \,\exp( \sum_{n=1}^\infty \left( 2GMn_d \mu^{2\epsilon} \right)^n (1-2\epsilon)  \int_{r'}^\infty   \frac{\dd{x}}{x^{1+n(1-2\epsilon)}} ).
    \label{eq:Green_fx_W_full}
\end{align}
where in the first equality, we plug in Eq.~\eqref{eq:Green_fx_u} and use the antisymmetry of the Wronskian.
The Green's function simplifies drastically since our source is localized at the origin, such that we only need to evaluate the Wronskian at $r'=0$. 
In this case, all the integrals in the exponent of above equation become scaleless, so we have
\begin{align}
    W[u^<_\ell, u^>_\ell] (0) =& W[u^-_\ell, u^+_\ell] (\infty)
    = 2i\omega
    \label{eq:Green_fx_W}
\end{align}
where the final equality follows from the asymptotic behavior~\eqref{eq:wave_FZ}. Using Eqs.~\eqref{eq:Green_fx_u} and~\eqref{eq:Green_fx_W}, we have the explit form of the Green's function in Eq.~\eqref{eq:Green_fx}.

After obtaining the Green's function, the waveform follows straightforwardly from
\ie 
    R_L (r) = \frac{1}{r^{1-\epsilon}} \int_0^\infty \dd{r'} G_\ell (r,r') \rho_L (r') = \frac{1}{r^{1-\epsilon}} \int_0^\infty \dd{r'} G_\ell (r,r') \left[ \frac{c_\ell}{\ell!} (-1)^{\ell+1} \mu^\epsilon r'^{1-\epsilon}  \sum_{\ell'}  Q^{L'} \int_{\Omega'} \hat{\vb*{r}}'_L \partial_{L'} \delta(\vb*{r}') \right] \,,
\fe
where we use the explicit form of $\rho_L$ in Eq.~\eqref{eq: def of source}. We can combine the radial and angular integration into a bulk integration
\ie
    R_L (r) &= \frac{1}{r^{1-\epsilon}} \frac{c_\ell}{\ell!} \mu^\epsilon  \sum_{\ell'} (-1)^{\ell+1}  Q^{L'} \int \dd[d-1]{\vb*{r}'} \frac{G_\ell (r,r')}{r'^{1-\epsilon}} \hat{\vb*{r}}'_L \partial_{L'} \delta(\vb*{r}')  \\
    &= \frac{1}{r^{1-\epsilon}} \frac{c_\ell}{\ell!} \mu^\epsilon  \sum_{\ell'} (-1)^{\ell+\ell'+1}  Q^{L'} \int \dd[d-1]{\vb*{r}'}  \delta(\vb*{r}') \partial_{L'} \left[  \frac{G_\ell (r,r')}{r'^{1-\epsilon}} \hat{\vb*{r}}'_L  \right] \,,
\fe
in which we perform the integration by part at the second line. To use the delta function, we have to evaluate the integrand at the origin $r' = 0$. Using Eqs.~\eqref{eq:Green_fx_u} and~\eqref{eq:Green_fx_W}, the integrand becomes
\begin{align}
    \partial_{L'} \left[ \frac{G_\ell(r,r')}{r'^{1-\epsilon}} \hat{\vb*{r}}'_L \right] =& u^+_\ell(r) \,\partial_{L'} \left[\frac{u^<_\ell(r')}{W[u^<_\ell, u^>_\ell] (r')}
    \frac{\vb*{r}'_L}{r'^{\ell+1-\epsilon}} \right]
    \label{eq:R_Green_1}
\end{align}
To find the $r'=0$ limit of the integrand, we need to identify the part that contains no integer power of $r'^\epsilon$. First, the inverse Wronskian $1/W[u^<_\ell, u^>_\ell] (0)=1/2i\omega$ starts as a constant at the origin. At finite $r'$, we can see from  Eq.~\eqref{eq:Green_fx_W_full} that the inverse Wronskian is corrected perturbatively by powers of $\left(GM/r'^{1-2\epsilon}\right)^n$. This correction only adds more positive powers of $r'^\epsilon$ which vanish at $r'=0$.
Second, we can expand $u^<_\ell$ in Eq.~\eqref{eq:Green_fx_u} in the near-zone basis. The function $\uirr$ always contains positive powers of $r^{\epsilon}$ even when the gravitational correction is included. On the other hand, the free part of $\ureg$, shown in Eq.~\eqref{eq:wave_NZ}, is the only term with $r^{-\epsilon}$ to cancel the $r'^{\ell+1-\epsilon}$ in the denominator in Eq.~\eqref{eq:R_Green_1}. 
This implies that we should only take $1/W[u^<_\ell, u^>_\ell] (r')=1/2i\omega$ and project $u^<_\ell$ to the $\ureg$ sector for the $r'=0$ limit which leads to
\begin{align}
    \partial_{L'} \left[ \frac{G_\ell(r,r')}{r'^{1-\epsilon}} \hat{\vb*{r}}'_L \right]\bigg\vert_{r'=0} =& \frac{u^+_\ell(r)}{2i\omega(W_{21}+c_\ell F_\ell W_{22})} \,\partial_{L'} \left[\frac{\ureg}{r'^{\ell+1-\epsilon}}
    \vb*{r}'_L \right]\bigg\vert_{r'=0} \nn \\
    =& \frac{u^+_\ell(r)\mu^{-\epsilon} \ell!}{2i\omega(W_{21}+c_\ell F_\ell W_{22})} \, \delta_{LL'}
    \label{eq:R_Green_2}
\end{align}
where we use Eqs~\eqref{eq:wave_NZ} and~\eqref{eq:Green_fx_u}, and  $\partial_{L'} \left(
    \vb*{r}'_L \right)=\ell! \delta_{L L'}$ in the second line.
Plugging Eq.~\eqref{eq:R_Green_1} to Eq.~\eqref{eq:R_Green_2}, we find
\ie
    R_L(r) =  \frac{ic_\ell}{2\omega( W_{21} + c_\ell F_\ell (\omega) W_{22} )} \frac{ \uout (r) }{ r^{1-\epsilon} } Q_L \,.
\fe
Combining the above with Eqs.~\eqref{eq: scalar wave spherical decomposition} and \eqref{eq: EFT far zone asymptotic bc}, we prove Eq.~\eqref{eq:phi_final} using the Green's function method.

\section{EFT Near-Zone Data}\label{app: EFT NZ data}

In the appendix, we discuss the renormalization of the near zone wavefunction. Once the near zone wavefunction has the UV divergence, we would need to introduce the contact term to cancel the divergence the get the renormalized wavefunction. We mainly follow the procedure in~\cite{Caron-Huot:2025tlq} by writing the bare boundary condition $B_{\rm reg}, B_{\rm irr}$ as a linear combination of the renormalized boundary condition $\bar B_{\rm reg}, \bar B_{\rm irr}$
\begin{equation}
\begin{pmatrix}
    B_{\rm reg} \\
    B_{\rm irr}
\end{pmatrix}
= 
\begin{pmatrix}
    Z_{11} & Z_{12} \\
    Z_{21} & Z_{22}
\end{pmatrix}
\begin{pmatrix}
    \bar B_{\rm reg} \\
    \bar B_{\rm irr}
\end{pmatrix}
~.
\end{equation}
From the RG equation
\begin{equation}
\mu \frac{d}{d\mu} 
\begin{pmatrix}
 \mu^{-\epsilon} B_{\rm reg} \\
 \mu^{\epsilon} B_{\rm irr}
\end{pmatrix}
=0 ~,
\end{equation}
we get that
\begin{equation}
\label{eq:B_RG}
\mu \frac{d}{d \mu} 
\begin{pmatrix}
    \bar B_{\rm reg} \\
    \bar B_{\rm irr}
\end{pmatrix}
= -
\begin{pmatrix}
    \gamma_{11}^{(\ell)} & \gamma_{12}^{(\ell)} \\
    \gamma_{21}^{(\ell)} & \gamma_{22}^{(\ell)}
\end{pmatrix}
\begin{pmatrix}
    \bar B_{\rm reg} \\
    \bar B_{\rm irr}
\end{pmatrix}
~, \quad \gamma^{(\ell)} = Z^{-1} \frac{d}{d\log \mu} Z + Z^{-1} 
\begin{pmatrix}
    - \epsilon & 0 \\
    0 & \epsilon
\end{pmatrix}
Z ~.
\end{equation}
In the situation where there is no source, we have 
\begin{equation}
\label{eq:tides_bc}
F_{\ell} = c_{\ell}^{-1} \frac{B_{\rm irr}}{B_{\rm reg}} = \frac{c_\ell^{-1} Z_{21} + Z_{22} \bar F_\ell}{Z_{11} + Z_{12} c_\ell \bar F_\ell} ~, \quad \bar{F}_{\ell} \equiv \frac{\bar B_{\rm irr}}{\bar B_{\rm reg}} ~.
\end{equation}
In this paper, we choose the physical minimal subtraction scheme on $F_\ell$ where $c_\ell^{-1} Z_{21}, Z_{22}, Z_{11}, Z_{12} c_\ell$ only contain singular in $\epsilon$ terms, i.e. $1/\epsilon, 1/\epsilon^2 \cdots$. It worths noting that this scheme is different than the $B_{\rm reg}, B_{\rm irr}$ minimal subtraction scheme used in~\cite{Caron-Huot:2025tlq} where they do not keep track $c_{\ell}$. Also, we choose to write everything in $G$ instead of $\bar{G} \equiv G M n_d$ as in~\cite{Caron-Huot:2025tlq}. As an example, for $\ell=1$ case until $\mathcal{O}(G^7)$, 
\begin{equation}
\begin{aligned}
    Z_{11} & = 1+ \frac{19}{\epsilon} x^2 + \Big(\frac{361}{1800 \epsilon ^2}+\frac{78037}{94500 \epsilon }\Big) x^4 + \Big(\frac{6859}{162000 \epsilon ^3}+\frac{1272703}{2835000 \epsilon ^2}+\frac{396789151}{111628125 \epsilon }\Big) x^6 ~, \\
    Z_{22} & = 1- \frac{19}{\epsilon} x^2 + \Big(\frac{361}{1800 \epsilon ^2}-\frac{78037}{94500 \epsilon }\Big) x^4 + \Big(-\frac{6859}{162000 \epsilon ^3}+\frac{1377703}{2835000 \epsilon ^2}-\frac{396789151}{111628125 \epsilon }\Big) x^6 ~, \\
    c_1 Z_{12} & = \frac{\omega^3}{\pi} \Big[ \frac{1}{12 \epsilon} x + \frac{2387}{5400 \epsilon} x^3 + \Big(\frac{361}{64800 \epsilon ^3}+\frac{119717}{1701000 \epsilon ^2}+\frac{306368569}{178605000 \epsilon }\Big) x^5 \\
    & \quad + \Big(\frac{1736471}{40824000 \pi  \epsilon ^3}+\frac{1411282069}{3572100000 \pi  \epsilon ^2}+\frac{279755900504}{35162859375 \pi  \epsilon }\Big)x^7\Big] ~,\\
    c_{1}^{-1} Z_{21} & = G^3 \pi \Big[ - \frac{4}{3\epsilon} x^2 + \Big(\frac{38}{135 \epsilon ^2}-\frac{19312}{2025 \epsilon }\Big) x^4\Big]  ~,
\end{aligned}
\end{equation}
where $x\equiv G M \omega$. The resulting $\gamma$ matrix in Eq.~\eqref{eq:B_RG} then gives
\begin{equation}
\gamma^{(\ell=0)} = 
\left(\begin{array}{cc}
\begin{aligned}
  &-\tfrac{22 x^2}{3} - \tfrac{9278 x^4}{135} - \tfrac{33555752 x^6}{42525}
\end{aligned}
&
\begin{aligned}
  & \omega \Big[-4 x - 52 x^3 - \tfrac{41110 x^5}{81} - \tfrac{1620387584 x^7}{273375}\Big]
\end{aligned}
\\[8ex]
G \Big[8 x^2 + \tfrac{304 x^4}{3} + \tfrac{798968 x^6}{675} \Big]
& 
\begin{aligned}
  &\tfrac{22 x^2}{3} + \tfrac{9278 x^4}{135} + \tfrac{33555752 x^6}{42525}
\end{aligned}
\end{array}\right)
\end{equation}

\begin{equation}
\gamma^{(\ell=1)} = 
\left(\begin{array}{cc}
\begin{aligned}
  &-\tfrac{38 x^2}{15} - \tfrac{156074 x^4}{23625} - \tfrac{1587156604 x^6}{37209375}
\end{aligned}
&
\begin{aligned}
  & \omega^3 \Big[-\tfrac{4 x}{9} - \tfrac{9548 x^3}{2025} - \tfrac{612737138 x^5}{22325625} - \tfrac{17904377632256 x^7}{105488578125}\Big]
\end{aligned}
\\[8ex]
G^3 \Big[8 x^2 + \frac{19312 x^4}{225} \Big]
& 
\begin{aligned}
  &\tfrac{38 x^2}{15} + \tfrac{156074 x^4}{23625} + \tfrac{1587156604 x^6}{37209375}
\end{aligned}
\end{array}\right)
\label{eq: gamma l1 appendix}
\end{equation}

\begin{equation}
\gamma^{(\ell=2)} = 
\left(\begin{array}{cc}
\begin{aligned}
  &-\tfrac{158 x^2}{105} - \tfrac{1416494 x^4}{1157625} - \tfrac{346394141024 x^6}{140390971875}
\end{aligned}
&
\begin{aligned}
  &\omega^5 \Big[-\tfrac{4 x}{225} - \tfrac{412124 x^3}{2480625} - \tfrac{18478507616 x^5}{27348890625} - \tfrac{241198148650093406 x^7}{109452311448046875}\Big]
\end{aligned}
\\[8ex]
G^5 \Big[\tfrac{32 x^2}{9} \Big]
& 
\begin{aligned}
  &\tfrac{158 x^2}{105} + \tfrac{1416494 x^4}{1157625} + \tfrac{346394141024 x^6}{140390971875}
\end{aligned}
\end{array}\right)
\end{equation}

\section{Black Hole Perturbation Theory Summary}
\label{app: BHPT match EFT}

In this appendix, we include some details used in the BHPT calculation. We use the MST method to solve the scalar wave equation \cite{Mino:1997bx,Mano:1996vt,Sasaki:2003xr}. After the decomposition (\ref{eq: scalar wave spherical decomposition}), one obtain one of the radial wave function as
\ie
R^C_\nu ( \omega r ) = & 2^{\nu} (  \omega r_s)^{-i \omega r_s } e^{-i \omega r } (\omega r)^{\nu + i\omega  r_s} \left( \frac{r}{r_s} - 1 \right)^{-i \omega r_s } \sum_{k=-\infty}^{\infty} \left( \sum_{n=-\infty}^{k} D^\nu_{n,k-n} \right)  (\omega r)^{k} ,\\
D^\nu_{n,j} = & (-1)^{n} (2i)^{\,n+j} \frac{ \Gamma(n+\nu+1 + 2iGM\omega) }{ \Gamma(2n + 2\nu + 2) } \frac{ (\nu + 1 - 2iGM\omega)_{n} }{ (\nu+1+ 2iGM\omega)_{n} } \frac{ (n+\nu+1 + 2iGM\omega)_{j} }{ (2n + 2\nu + 2)_j \, j! } a^\nu_n, 
\label{eq: Rc def}
\fe
where $\nu$ is the renormalized angular momentum and $a^\nu_n$ are the MST coefficients. The other linear independent solution is $R^C_{-\nu - 1}$. To relate these solutions to near zone basis, we need to consider the limit $\omega r_s \ll \omega r \ll 1$ as explained in the main text. Hence, we perform Taylor expansion for the functions $e^{i\omega r}$ and $(1-r_s/r)^{-i\omega r_s}$ in (\ref{eq: Rc def}). Rearranging the summations, we arrive at 
\begin{align}
    R^C_\nu (r) &= (2\omega r)^\nu \sum_{j = -\infty}^\infty \tilde{D}^\nu_j (\omega r)^j  \;\;,\;\; \tilde{D}^\nu_j = \sum_{m=j}^\infty (-\omega r_s)^{m-j} \binom{-i\omega r_s}{m-j} \sum_{k=-\infty}^m \frac{(-i)^{m-k} }{(m-k)!} \sum_{n=-\infty}^k D^\nu_{n,k-n} \label{eq: Dtilde def}
\end{align}
Since the outermost summation on $m$ in $\tilde{D}^\nu_j$ is organized in powers of $\omega r_s$, it can be truncated at a fixed order in $G$. Moreover, because the innermost summation on $n$ involves $D^\nu_{n,k-n}$, which contains the MST coefficients, it is bounded from below and can be truncated accordingly. As a result, when we compute $\tilde{D}_j^\nu$ perturbatively to a fixed order in $G$, the ranges of summations in Eq.~\eqref{eq: Dtilde def} become finite.

The other solution basis of BHPT is defined as $(R^\tx{up}_\nu (r) ,R^\tx{down}_\nu (r))$, 
whose asymptotic behaviors are
\begin{align}
    R^\tx{up}_\nu (r \to \infty ) &= A^\tx{up}_\nu \frac{ e^{+i\omega(r+ 2GM \log(\omega r))} }{ \omega r}  \label{eq: R up asymptotic}\\
    R^\tx{down}_\nu (r \to \infty ) &= A^\tx{down}_\nu  \frac{ e^{-i\omega ( r+ 2GM \log(\omega r))} }{ \omega r } ,
    \label{eq: R down asymptotic}
\end{align}
where the coefficients $A^\tx{up}_\nu,A^\tx{down}_\nu$ are given by
\begin{align}
    A^\tx{up}_\nu &= 2^{-1+2iGM\omega} e^{-\frac{i\pi}{2} (\nu+1) } e^{-\pi GM\omega} \sum_{n=-\infty}^\infty (-1)^n \frac{(\nu+1-2iGM \omega)_n}{(\nu+1+2iGM \omega)_n} a^\nu_n \label{eq: Aup def} \\
    A^\tx{down}_\nu &= 2^{-1-2iGM\omega} e^{\frac{i\pi}{2} (\nu+1) } e^{-\pi G M\omega } \frac{\Gamma(\nu+1+2iGM\omega)}{\Gamma(\nu+1-2iGM\omega)} \sum_{n=-\infty}^\infty a^\nu_n .
    \label{eq: Adown def}
\end{align}
The two solution basis are related by
\begin{align}
    R_\nu^C (r) &= R_\nu^\tx{down} (r) + R_\nu^\tx{up} (r)  ,
    \label{eq: R up and down def 1}\\
    R_{-\nu-1}^C (r) &= \frac{ A^\tx{down}_{-\nu-1} }{ A^\tx{down}_\nu } R^\tx{down}_\nu (r) + \frac{ A^\tx{up}_{-\nu-1} }{ A^\tx{up}_\nu } R^\tx{up}_\nu (r).
    \label{eq: R up and down def 2}
\end{align}
Therefore, we can identify the connection matrix defined in Eq. (\ref{eq:W_BHPT}) as
\begin{align}
    \vb*{W}^\tx{BHPT} = \begin{pmatrix}
        1 & A^\tx{up}_{-\nu-1} / A^\tx{up}_\nu  \\
        1 &  A^\tx{down}_{-\nu-1} / A^\tx{down}_\nu 
    \end{pmatrix} .
    \label{eq: Wronskian BHPT}
\end{align}
Finally, comparing with asymptotic behaviors of the far zone basis (\ref{eq: EFT far zone asymptotic bc}), we find
\begin{align}
    \vb*{T}^\tx{FZ} = 
    \begin{pmatrix}
        \dfrac{ \omega }{ i^{\ell+1} A^\tx{up}_\nu}  e^{-i\delta_\tx{IR} } & 0 \\
        0 & \dfrac{ i^{\ell +1} \omega}{   A^\tx{down}_\nu} e^{ i\delta_\tx{IR} }
    \end{pmatrix}  \quad \tx{with} \;\; \delta_\tx{IR} = GM\omega \log(\frac{\omega^2}{\bar{\mu}^2_\tx{IR}}) \label{eq: Far Transition matrix}\,.
\end{align}

\section{Renormalization Group Equation of Tides and Radiative Multiples} \label{app: anomalous dimension and RG}
In this appendix, we discuss the details of the renormalization group equations of tides and radiative multiples. We will first derive the equation and then give the solution and interpretations. 

First of all, we note that the near zone wavefunction can be written as
\begin{equation}
    \bar u_{\ell m} = \bar B_{\rm reg} \bar u_{\rm reg} + \bar B_{\rm irr} \bar u_{\rm irr} = 
    \begin{pmatrix}
        \bar u_{\rm reg} & \bar u_{\rm irr}
    \end{pmatrix}
    \begin{pmatrix}
        \bar B_{\rm reg} \\
        \bar B_{\rm irr}
    \end{pmatrix}
    ~
\end{equation}
which is independent of the renormalization scale $\mu$. Therefore, we get
\begin{equation}
    \mu \frac{d}{d\mu} 
    \begin{pmatrix}
        \bar u_{\rm reg} \\
        \bar u_{\rm irr}
    \end{pmatrix}
    = \vb*{\gamma}^{\rm T}
    \begin{pmatrix}
        \bar u_{\rm reg} \\
        \bar u_{\rm irr}
    \end{pmatrix}
    ~.
\end{equation}
From the connection matrix formula with Wronkian $W$
\begin{equation}
    (\ureg,\uirr)= (\uout,\uin) \,\vb*{W}
\end{equation}
and the fact that the far-zone base function $\uout,\uin$ does not have UV divergence, we arrive at
\begin{align}
\mu \frac{d}{d\mu} \vb*{W} = \vb*{W} \vb*{\gamma}.
\end{align}
By using the identity $\det \vb*{W}  = 1/(2i\omega)$, we find
\begin{equation}
    \gamma^{(\ell)}_{11} = - \gamma_{22}^{(\ell)} ~.
\end{equation}
The above discussion is purely at the level of differential equation and universally applies to all the physically relevant process. If we consider the scattering process where there is no source and the ratio $B_{\rm irr}/B_{\rm reg}$ is fixed by tidal response function $F_{\ell}$, i.e. $c_{\ell} \bar F_{\ell} = \bar B_{\rm irr} / \bar B_{\rm reg}$, we can derive the RG equation for $\bar F$ by requiring the scattering amplitude in Eq.~\eqref{eq:S_matrix} to be independent of $\mu$
\begin{equation}
\label{eq:app_F_RG}
\mu \frac{d S_{\ell}}{d\mu} = 0 ~ \quad  \Rightarrow  \quad     \mu \dv{}{\mu}\bar{F}_\ell = - \frac{\gamma_{21}}{c_\ell} + (\gamma_{11} - \gamma_{22}) \bar{F}_\ell + c_\ell \gamma_{12} \bar{F}^2_\ell ~.
\end{equation}
If we consider the emission process with boundary condition Eq.~\eqref{eq: near zone boundary condition} and the outgoing boundary condition at infinity, we can derive the RG equation for $\bar Q$ by setting the waveform at infinity to be independent of $\mu$
\begin{equation}
\label{eq:app_Q_RG}
    \mu \frac{d R_L}{d\mu} = 0 ~ \quad  \Rightarrow  \quad   \mu\dv{}{\mu} \bar{Q}_L 
= \left( \gamma_{11} + c_\ell \gamma_{12}\bar{F}_\ell \right) \bar{Q}_L ~.
\end{equation}

Through the perturbative calculation, we also notice that
\begin{equation}
    \tx{ The eigenvalues of}\; \gamma \;\tx{are}\; \pm( \ell - \nu )  ~.
\end{equation}
The proof is very straightforward. Let us consider a change of basis in the two dimensional solution space of the EFT wave equation
\begin{equation}
    \begin{pmatrix}
         \bar u'_{\rm reg} \\
         \bar u'_{\rm irr}
    \end{pmatrix}
    = O 
    \begin{pmatrix}
        \bar u_{\rm reg} \\
        \bar u_{\rm irr}
    \end{pmatrix} 
    ~.
\end{equation}
In the new basis $\bar u'_{\rm reg}, \bar u'_{\rm irr}$, the anomalous dimension matrix undergoes the similarity transformation
\begin{equation}
    \mu \frac{d}{d\mu}
    \begin{pmatrix}
        \bar u'_{\rm reg} \\
        \bar u'_{\rm irr}
    \end{pmatrix}
    = O \gamma^{\rm T} O^{-1}
    \begin{pmatrix}
        \bar u'_{\rm reg} \\
        \bar u'_{\rm irr}
    \end{pmatrix}
    ~,
\end{equation}
which preserves the trace and eigenvalue.
Therefore, there exists a basis where the anomalous dimension matrix is diagonal. Remarkably, this basis exists in BHPT. To see that, we first take the two solutions $R_\nu^C, R_{-\nu-1}^C$ in BHPT. These two solutions has the following asymptotic scaling behavior
\begin{align}
    R^C_\nu &= e^{(\nu-\ell) \log(2\omega r)} \Big( (2\omega r)^\ell + \cdots + (2\omega r)^{-\ell-1} + \cdots \Big) ~,
     \\
    R^C_{-\nu-1} &= e^{-(\nu-\ell) \log(2\omega r)} \Big(  (2\omega r)^\ell + \cdots + (2\omega r)^{-\ell-1} + \cdots \Big) ~.
\end{align}
This solution has no manifest $\mu$ dependence. However, to match with the EFT solution, we can manually introduce the $\mu$ dependence using the trick
\begin{equation}
    \log(\omega r) = \log\Big(\frac{\omega}{\mu}\Big) + \log(\mu r) ~.
\end{equation}
This motivates us to define the $\mu$-dependent BHPT solution
\begin{align}
    \bar R_{\nu}^C = e^{(\nu-\ell) \log(2 \mu r)}((2 \omega r)^\ell + \cdots + (2 \omega r)^{-\ell-1} + \cdots) ~, \\
    \bar R_{-\nu-1}^C = e^{-(\nu-\ell) \log(2 \mu r)}((2 \omega r)^\ell + \cdots + (2 \omega r)^{-\ell-1} + \cdots) ~.
\end{align}
In this basis, it is now manifest that the eigenvalue of the anomalous dimension matrix is $\pm (\ell-\nu)$.

\begin{figure}[t]
\centering
\includegraphics[width=0.5\textwidth]{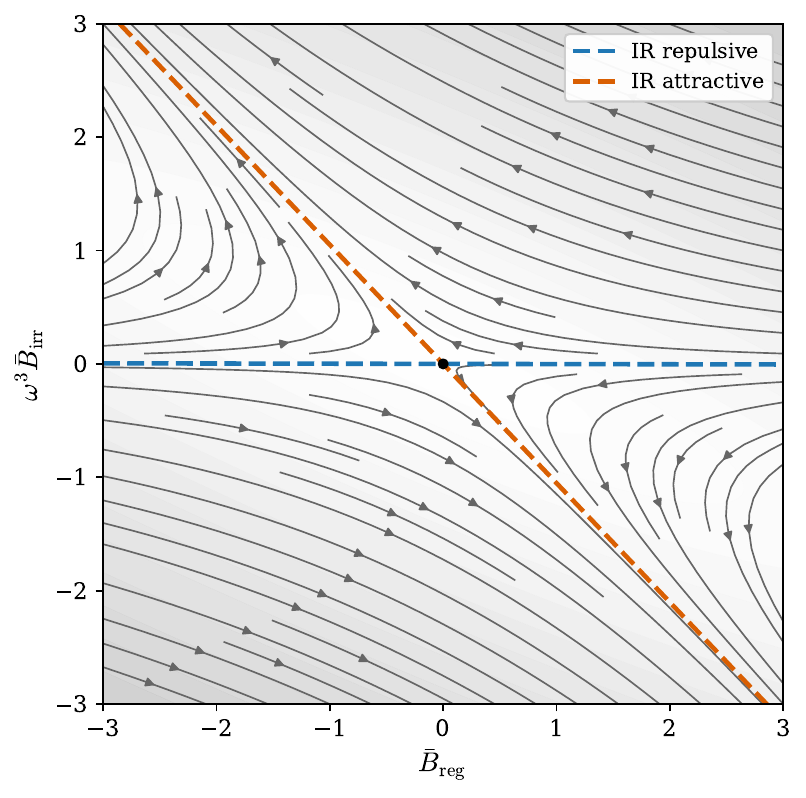}
\caption{Numerical RG flow in the $(\bar B_{\rm reg},\omega^3\bar B_{\rm irr})$ plane for $\ell=1$ at $x=GM\omega=0.1$, using the truncated anomalous dimension matrix in Eq.~\eqref{eq: gamma l1 appendix}. The streamlines are obtained after rewriting the linear system in the rescaled basis $(\bar B_{\rm reg},\omega^3\bar B_{\rm irr})$, so that the flow depends only on $x$. The flatter dashed ray is the IR-repulsive eigendirection, associated with $\omega^3 c_1\bar F_{1,+}\simeq -1.70\times 10^{-3}$, while the steeper dashed ray is the IR-attractive eigendirection, associated with $\omega^3 c_1\bar F_{1,-}\simeq -1.05$.}
\label{fig:RG-flow-B-plane}
\end{figure}

We now solve the RG equations exactly at fixed perturbative order, where $\gamma_{ij}(\omega)$ are $\mu$ independent. We choose $\mu_0$ as our initial energy scale and impose the initial conditions
\begin{align}
\bar{F}_\ell(\omega,\mu_0) = \bar{F}_{\ell,0}(\omega), \qquad
\bar{Q}_L(\omega,\mu_0) = \bar{Q}_{L,0}(\omega).
\end{align}
We are now going to first solve Eq.~\eqref{eq:app_F_RG} and Eq.~\eqref{eq:app_Q_RG} with generic $\gamma$ matrix. Define the discriminant
\begin{align}
\Lambda_\ell(\omega) \equiv
\sqrt{\bigl(\gamma_{11}(\omega)-\gamma_{22}(\omega)\bigr)^2 + 4\,\gamma_{12}(\omega)\gamma_{21}(\omega)}
\end{align}
and the two fixed points
\begin{align}
\bar{F}_{\ell,\mp}(\omega) \equiv
\frac{\bigl(\gamma_{11}(\omega)-\gamma_{22}(\omega)\bigr) \mp \Lambda_\ell(\omega)}{2\,c_\ell (-\gamma_{12}(\omega))}.
\end{align}
We find that these two fixed points actually corresponds to the eigendirections of the RG flow in $(\bar B_{\rm reg}, \bar B_{\rm irr})$ plane. Since $\gamma_{12}^{(\ell=1)}$ and $\gamma_{21}^{(\ell=1)}$ carry explicit factors $\omega^3$ and $G^3$, the raw $(\bar B_{\rm reg}, \bar B_{\rm irr})$ plane is not the most convenient one for a numerical illustration. It is cleaner to work in the dimensionless plane $(\bar B_{\rm reg},\omega^3 \bar B_{\rm irr})$, where the flow depends only on $x=GM\omega$. For illustration, taking the $\ell=1$ anomalous dimension matrix in Eq.~\eqref{eq: gamma l1 appendix} and choosing $x=GM\omega=0.1$, we find
\begin{align}
\omega^3 c_1 \bar F_{1,+} \simeq -1.70 \times 10^{-3},
\qquad
\omega^3 c_1 \bar F_{1,-} \simeq -1.05,
\end{align}
so the flatter eigendirection is UV repulsive while the steeper one is UV attractive, as shown explicitly in Fig.~\ref{fig:RG-flow-B-plane}. We can then rewrite the RG equation as
\begin{align}
\dv{}{\log\mu}\bar{F}_\ell
=
c_\ell \gamma_{12}(\omega)\,
\bigl(\bar{F}_\ell-\bar{F}_{\ell,+}\bigr)
\bigl(\bar{F}_\ell-\bar{F}_{\ell,-}\bigr),
\qquad
c_\ell (-\gamma_{12}(\omega))\,
\bigl(\bar{F}_{\ell,+}-\bar{F}_{\ell,-}\bigr)
=
\Lambda_\ell(\omega).
\end{align}
The exact solution is therefore
\begin{align}
\frac{\bar{F}_\ell(\omega,\mu)-\bar{F}_{\ell,-}(\omega)}{\bar{F}_\ell(\omega,\mu)-\bar{F}_{\ell,+}(\omega)}
=
\frac{\bar{F}_{\ell,0}(\omega)-\bar{F}_{\ell,-}(\omega)}{\bar{F}_{\ell,0}(\omega)-\bar{F}_{\ell,+}(\omega)}
\left( \frac{\mu}{\mu_0} \right)^{\Lambda_\ell(\omega)} \label{eq: exact RG solution F implicit}
\end{align}
or equivalently
\begin{align}
\bar{F}_\ell(\omega,\mu)
=
\bar{F}_{\ell,+}(\omega)
+
\frac{\bigl(\bar{F}_{\ell,0}(\omega)-\bar{F}_{\ell,+}(\omega)\bigr)\bigl(\bar{F}_{\ell,-}(\omega)-\bar{F}_{\ell,+}(\omega)\bigr)\left( \frac{\mu}{\mu_0} \right)^{-\Lambda_\ell(\omega)}}{\bar{F}_{\ell,-}(\omega)-\bar{F}_{\ell,0}(\omega)+\bigl(\bar{F}_{\ell,0}(\omega)-\bar{F}_{\ell,+}(\omega)\bigr)\left( \frac{\mu}{\mu_0} \right)^{-\Lambda_\ell(\omega)}}.
\label{eq: exact RG solution F first}
\end{align}

Once $\bar{F}_\ell$ is known, the exact solution of $\bar{Q}_L$ is
\begin{align}
\bar{Q}_L(\omega,\mu)
=
\bar{Q}_{L,0}(\omega)\,
\frac{\bar{F}_{\ell,-}(\omega)-\bar{F}_{\ell,+}(\omega)}{\bar{F}_{\ell,-}(\omega)-\bar{F}_{\ell,0}(\omega)+\bigl(\bar{F}_{\ell,0}(\omega)-\bar{F}_{\ell,+}(\omega)\bigr)\left( \frac{\mu}{\mu_0} \right)^{-\Lambda_\ell(\omega)}}
\left( \frac{\mu}{\mu_0} \right)^{\gamma_{11}(\omega)+c_\ell \gamma_{12}(\omega)\bar{F}_{\ell,+}(\omega)}. \label{eq: exact RG solution Q first}
\end{align}
Using
\begin{align}
c_\ell (-\gamma_{12}(\omega)) \bar{F}_{\ell,+}(\omega)
=
\frac{\bigl(\gamma_{11}(\omega)-\gamma_{22}(\omega)\bigr)+\Lambda_\ell(\omega)}{2},
\end{align}
the exponent can be rewritten as
\begin{align}
\gamma_{11}(\omega)+c_\ell \gamma_{12}(\omega)\bar{F}_{\ell,+}(\omega)
=
\frac{\gamma_{11}(\omega)+\gamma_{22}(\omega)-\Lambda_\ell(\omega)}{2}.
\end{align}
Hence
\begin{align}
\label{eq:Q_RG_sol_org}
\bar{Q}_L(\omega,\mu)
=
\bar{Q}_{L,0}(\omega)\,
\frac{\bar{F}_{\ell,-}(\omega)-\bar{F}_{\ell,+}(\omega)}{\bar{F}_{\ell,-}(\omega)-\bar{F}_{\ell,0}(\omega)+\bigl(\bar{F}_{\ell,0}(\omega)-\bar{F}_{\ell,+}(\omega)\bigr)\left( \frac{\mu}{\mu_0} \right)^{-\Lambda_\ell(\omega)}}
\left( \frac{\mu}{\mu_0} \right)^{\frac{\gamma_{11}(\omega)+\gamma_{22}(\omega)-\Lambda_\ell(\omega)}{2}}. 
\end{align}
Now, we simplify Eq.~\eqref{eq: exact RG solution F implicit} and Eq.~\eqref{eq:Q_RG_sol_org} using the two properties: ${\rm Tr} \gamma = 0$ and $\det \gamma = - (\nu(\omega)-\ell)^2$. In particular, we find that the discriminant can be simplified to
\begin{equation}
    \Lambda_{\ell}(\omega) = \sqrt{(\gamma_{11}-\gamma_{22})^2 + 4\,\gamma_{12}\gamma_{21}} = 2\sqrt{- \gamma_{11} \gamma_{22} + \gamma_{12} \gamma_{21}} = 2 (\ell-\nu(\omega)) ~,
\end{equation}
where $\nu$ is the fixed-$\ell$ renormalized angular momentum in BHPT. The above formula allows us to simplify the solution to the RG equation to be
\begin{align}
\bar{Q}_L(\omega,\mu)
=
\bar{Q}_{L,0}(\omega)\,
\frac{\bar{F}_{\ell,-}(\omega)-\bar{F}_{\ell,+}(\omega)}{\bar{F}_{\ell,-}(\omega)-\bar{F}_{\ell,0}(\omega)+\bigl(\bar{F}_{\ell,0}(\omega)-\bar{F}_{\ell,+}(\omega)\bigr)\left( \frac{\mu}{\mu_0} \right)^{2(\nu(\omega)-\ell)}}
\left( \frac{\mu}{\mu_0} \right)^{(\nu(\omega)-\ell)}. \label{eq: exact RG solution Q}
\end{align}
and 
\begin{align}
\bar{F}_\ell(\omega,\mu)
=
\bar{F}_{\ell,+}(\omega)
+
\frac{\bigl(\bar{F}_{\ell,0}(\omega)-\bar{F}_{\ell,+}(\omega)\bigr)\bigl(\bar{F}_{\ell,-}(\omega)-\bar{F}_{\ell,+}(\omega)\bigr)\left( \frac{\mu}{\mu_0} \right)^{2(\nu(\omega)-\ell)}}{\bar{F}_{\ell,-}(\omega)-\bar{F}_{\ell,0}(\omega)+\bigl(\bar{F}_{\ell,0}(\omega)-\bar{F}_{\ell,+}(\omega)\bigr)\left( \frac{\mu}{\mu_0} \right)^{2(\nu(\omega)-\ell)}}. 
\end{align}
These two equations allow us to get better resummation proposal.

\section{Perturbative Results on the Sommerfeld Factors}\label{app: sommerfeld results}
In this appendix, we give more details on the calculation of the gravitational Sommerfeld factor. First of all,  we emphasize that the physical outgoing waveform 
\begin{equation}
\label{eqapp:renom_asymp}
R_L\big|_{r\to\infty}
=
-\frac{c_\ell}{W_{21}(\omega,\mu)+c_\ell \bar F_\ell(\omega,\mu,\mu_0)\,W_{22}(\omega,\mu)}
\frac{\bar Q_L(\omega,\mu,\mu_0)}{2 i \omega}\,
\frac{e^{i\omega r}}{r}\, .
\end{equation}
is independent of the renormalization scale $\mu$, but depends on the physical scale of the UV theory $\mu_0$. It is often convenient to express the result by choosing $\mu=\mu_0$, so that the waveform can be written as
\begin{align}
R_L\big|_{r\to\infty}
=
-\frac{c_\ell}{W_{21}(\omega,\mu_0)+c_\ell \bar F_{\ell,0}\,W_{22}(\omega,\mu_0)}
\frac{\bar Q_{L,0}(\omega)}{2 i \omega}\,
\frac{e^{i\omega r}}{r}\, .
\label{eq:Rout_mu0}
\end{align}
From this equation, we define the Sommerfeld factor as
\begin{align}
\mathcal{S}_\ell(\omega;\mu_0)
\equiv
\frac{R_{L}|_{r\to\infty}}{R_{L}|_{G=0,F=0}}
=
\frac{(2\ell+1)!!}
{2\omega^{\ell+1}\Bigl(W_{21}(\omega,\mu_0)+c_\ell \bar F_{\ell,0}(\omega)\,W_{22}(\omega,\mu_0)\Bigr)} \, .
\label{eq:Sommerfeld_mu0}
\end{align}
In this paper, we are going to report the Sommerfeld factor in a regime where $\bar F_{\ell,0}(\omega)$ takes the Taylor-expanded form
\begin{equation}
    \bar F_{\ell,0}(\omega) = (G M)^{2\ell+1} \sum_{n=0}^\infty c_{\ell,n}(\mu_0) (i G M \omega)^n.
    \label{eq:tidal_rescaled}
\end{equation}
In the following, we present the Sommerfeld factor until $\mathcal{O}(G^{10})$ using the definitions
\begin{align}
    x \equiv GM\omega, \qquad
    L \equiv \log\left(\frac{4 \omega^2}{\bar{\mu}^2}\right), \qquad
    L_{\tx{IR}} \equiv \log\left(\frac{4 \omega^2}{\bar{\mu}_{\tx{IR}}^2}\right),
\end{align}
with $\bar{\mu}^2 = 4\pi e^{-\gamma_E} \mu_0^2, \bar{\mu}_{\mathrm{IR}}^2=\mu_{\mathrm{IR}}^2 4 \pi e^{\gamma_E-1}$. We calculate the magnitude $|\mathcal{S}_{\ell}|$ and phase $\tx{Arg} \mathcal{S}_{\ell}$ of the Sommerfeld enhancement factor up to $\mathcal{O}(x^{10})$ in the minimal subtraction scheme and the parametrization in Eq.~\eqref{eq:tidal_rescaled}. 
For $\ell=0,1,2$, we present the explicit results up to $\mathcal{O}(x^{5})$.

\paragraph{Magnitude for ${\ell=0}$:}
\begin{align}
|\mathcal{S}_{\ell=0}| = &1 
+ x \pi  \nonumber \\
&+ x^2 \Biggl[
\tfrac{67}{6} + \tfrac{\pi^2}{6} + \tfrac{c_{0,0}}{2\pi} - \tfrac{c_{0,0}^2}{32\pi^2} - \tfrac{c_{0,1}}{4\pi}
+ L \left( -\tfrac{11}{3} - \tfrac{c_{0,0}}{2\pi} \right)
\Biggr] \nonumber  \\
&+ x^3 \Biggl[
\tfrac{43\pi}{6} - \tfrac{\pi^3}{6} - \tfrac{4}{3} c_{0,0} - \tfrac{5 c_{0,0}^2}{32\pi} - \tfrac{3}{4} c_{0,1}
+ L \left( -\tfrac{11\pi}{3} - \tfrac{1}{2} c_{0,0} \right)
\Biggr] \nonumber \\
&+ x^4 \Biggl[
\tfrac{629411}{3240} - \tfrac{57\pi^2}{4} - \tfrac{\pi^4}{40} - \tfrac{88}{3} \zeta(3)
- \tfrac{3 c_{0,0}^3}{64\pi^3} + \tfrac{3 c_{0,0}^4}{2048\pi^4}
- \tfrac{c_{0,2}}{2\pi} + \tfrac{c_{0,3}}{4\pi}
+ \left( -\tfrac{67}{8\pi} - \tfrac{7\pi}{8} \right) c_{0,1} \nonumber  \\
&\quad + \tfrac{c_{0,1}^2}{16\pi^2} 
+ c_{0,0}^2 \left( -\tfrac{65}{192} - \tfrac{287}{192\pi^2} + \tfrac{3 c_{0,1}}{128\pi^3} \right)
+ c_{0,0} \left( \tfrac{151}{8\pi} - \tfrac{173\pi}{36} - \tfrac{3 c_{0,1}}{8\pi^2} + \tfrac{c_{0,2}}{16\pi^2} - \tfrac{4 \zeta(3)}{\pi} \right) \nonumber  \\
&\quad
+ L \Biggl(
-\tfrac{22493}{270} - \tfrac{11\pi^2}{18} + \tfrac{7 c_{0,0}^2}{96\pi^2}
+ \tfrac{3 c_{0,0}^3}{64\pi^3} + \tfrac{11 c_{0,1}}{4\pi}
+ c_{0,0} \left( -\tfrac{199}{12\pi} - \tfrac{\pi}{12} + \tfrac{3 c_{0,1}}{8\pi^2} \right)
+ \tfrac{c_{0,2}}{2\pi}
\Biggr) \nonumber  \\
&\quad
+ L^2 \left(
\tfrac{193}{18} + \tfrac{11 c_{0,0}}{3\pi} + \tfrac{c_{0,0}^2}{4\pi^2}
\right)
\Biggr] \nonumber \\
&+ x^5 \Biggl[
\tfrac{31091\pi}{3240} - \tfrac{719\pi^3}{36} + \tfrac{19\pi^5}{360}
- \tfrac{c_{0,0}^3}{16\pi^2} + \tfrac{27 c_{0,0}^4}{2048\pi^3}
- \tfrac{88}{3}\pi \zeta(3)
+ \left( -\tfrac{177}{8} - \tfrac{3\pi^2}{8} \right) c_{0,1}
+ \tfrac{5 c_{0,1}^2}{16\pi} \nonumber \\
&\quad
+ c_{0,0}^2 \left( -\tfrac{1859}{192\pi} - \tfrac{25\pi}{64} + \tfrac{21 c_{0,1}}{128\pi^2} \right)
+ c_{0,0} \left( -\tfrac{70981}{1080} - \tfrac{53\pi^2}{9} + \tfrac{c_{0,1}}{4\pi} + \tfrac{5 c_{0,2}}{16\pi} - 4 \zeta(3) \right)
+ \tfrac{4}{3} c_{0,2} + \tfrac{3}{4} c_{0,3} \nonumber \\
&\quad
+ L \Biggl(
-\tfrac{10613\pi}{270} + \tfrac{11\pi^3}{18}
+ \tfrac{403 c_{0,0}^2}{96\pi} + \tfrac{15 c_{0,0}^3}{64\pi^2}
+ \tfrac{33}{4} c_{0,1}
+ c_{0,0} \left( \tfrac{115}{12} + \tfrac{\pi^2}{12} + \tfrac{9 c_{0,1}}{8\pi} \right)
+ \tfrac{1}{2} c_{0,2}
\Biggr) \nonumber \\
&\quad
+ L^2 \left(
\tfrac{193\pi}{18} + \tfrac{11}{3} c_{0,0} + \tfrac{c_{0,0}^2}{4\pi}
\right)
\Biggr] + \mathcal{O}(x^6)\,.
\end{align}

\paragraph{Phase for ${\ell=0}$:}
\begin{align}
\tx{Arg} \mathcal{S}_{\ell=0} &= x \left[
-1 + 2 \gamma_E + L_{\tx{IR}} + \tfrac{c_{0,0}}{4\pi}
\right] + x^2 \left[
\tfrac{11\pi}{3} + \tfrac{1}{2} c_{0,0}
\right] \nonumber + x^3 \Bigg[
\tfrac{50}{3} + \tfrac{22\pi^2}{9} - \tfrac{8}{3}\zeta(3)
+ \tfrac{c_{0,0}^2}{8\pi^2} - \tfrac{c_{0,0}^3}{192\pi^3}
+ \tfrac{c_{0,1}}{2\pi} - \tfrac{c_{0,2}}{4\pi} \nonumber \\
&\quad
+ c_{0,0} \left( \tfrac{67}{12\pi} + \tfrac{\pi}{3} - \tfrac{c_{0,1}}{16\pi^2} \right) + L \left(
-4 - \tfrac{11 c_{0,0}}{6\pi} - \tfrac{c_{0,0}^2}{8\pi^2} - \tfrac{c_{0,1}}{2\pi}
\right)
\Bigg] \nonumber \\
&+ x^4 \Biggl[
\tfrac{5719\pi}{135}
- \tfrac{5 c_{0,0}^2}{24\pi} - \tfrac{c_{0,0}^3}{32\pi^2}
- \tfrac{11}{6} c_{0,1} - \tfrac{1}{2} c_{0,2} + c_{0,0} \left( \tfrac{55}{6} - \tfrac{c_{0,1}}{4\pi} \right) + L \left(
-8\pi - \tfrac{11}{3} c_{0,0} - \tfrac{c_{0,0}^2}{4\pi}
\right)
\Biggr] \nonumber \\
&+ x^5 \Biggl[
\tfrac{51728}{135} + \tfrac{5498\pi^2}{405} - \tfrac{88\pi^4}{135}
+ \tfrac{104}{9}\zeta(3) + \tfrac{32}{5}\zeta(5)
- \tfrac{c_{0,0}^4}{128\pi^4} + \tfrac{c_{0,0}^5}{5120\pi^5}
- \tfrac{c_{0,1}^2}{8\pi^2} \nonumber \\
&\quad
+ c_{0,0}^3 \left( -\tfrac{55}{192\pi^3} - \tfrac{1}{12\pi} + \tfrac{c_{0,1}}{256\pi^4} \right) + c_{0,0}^2 \left( -\tfrac{109}{72} + \tfrac{587}{96\pi^2} - \tfrac{3 c_{0,1}}{32\pi^3} + \tfrac{c_{0,2}}{64\pi^3} - \tfrac{\zeta(3)}{\pi^2} \right) \nonumber \\
&\quad
+ c_{0,0} \Biggl(
\tfrac{103927}{810\pi} - \tfrac{47\pi}{18} - \tfrac{4\pi^3}{45}
+ \left( -\tfrac{5}{12} - \tfrac{61}{24\pi^2} \right) c_{0,1}
+ \tfrac{c_{0,1}^2}{64\pi^3}
- \tfrac{c_{0,2}}{4\pi^2}
+ \tfrac{c_{0,3}}{16\pi^2}
- \tfrac{44}{3\pi}\zeta(3)
\Biggr) + \mathcal{O}(x^6)\,.
\end{align}

\paragraph{Magnitude for ${\ell=1}$:}
\begin{align}
|\mathcal{S}_{\ell = 1}| &= 1 + x  \pi + x^2 \left[
\tfrac{413}{50} + \tfrac{\pi^2}{6} - \tfrac{19}{15} L
\right] + x^3 \left[
\tfrac{413\pi}{50} - \tfrac{\pi^3}{6}
- \tfrac{19\pi}{15} L
\right] \nonumber \\
&+ x^4 \Biggl[
\tfrac{1292307799}{19845000} - \tfrac{41\pi^2}{180} - \tfrac{\pi^4}{40}
- \tfrac{152}{15}\zeta(3)
+ \tfrac{7 c_{1,0}}{18\pi} - \tfrac{c_{1,1}}{12\pi} + L \left(
-\tfrac{130087}{9450} - \tfrac{19\pi^2}{90} - \tfrac{c_{1,0}}{6\pi} \right)
+ \tfrac{361}{450} L^2 \Biggr] \nonumber \\
&+ x^5 \Biggl[
\tfrac{1283487799\pi}{19845000} - \tfrac{2683\pi^3}{900} + \tfrac{19\pi^5}{360}
- \tfrac{152}{15}\pi \zeta(3)
+ \tfrac{8}{45} c_{1,0} - \tfrac{1}{4} c_{1,1} + L \left(
-\tfrac{130087\pi}{9450} + \tfrac{19\pi^3}{90} - \tfrac{1}{6} c_{1,0}
\right)
+ \tfrac{361\pi}{450} L^2 \Biggr] + \mathcal{O}(x^6)\,,
\end{align}

\paragraph{Phase for ${\ell=1}$:}
\begin{align}
\tx{Arg} \mathcal{S}_{\ell=1} &= x \Bigl[
-3 + 2 \gamma_E + L_{\tx{IR}}
\Bigr] + \tfrac{19\pi}{15} x^2 + x^3 \left[
\tfrac{38\pi^2}{45} - \tfrac{8}{3}\zeta(3)
+ \tfrac{c_{1,0}}{12\pi}
\right] + x^4 \left[
\tfrac{78037\pi}{23625} + \tfrac{1}{6} c_{1,0}
\right] \nonumber \\
&+ x^5 \Biggl[
\tfrac{2822}{675} + \tfrac{156074\pi^2}{70875} - \tfrac{152\pi^4}{675}
+ \tfrac{2888}{225}\zeta(3) + \tfrac{32}{5}\zeta(5)
+ \tfrac{413 c_{1,0}}{300\pi} + \tfrac{\pi}{9} c_{1,0}
+ \tfrac{7 c_{1,1}}{18\pi} - \tfrac{c_{1,2}}{12\pi} \nonumber \\
&\quad
+ L \left(
-\tfrac{4}{9} - \tfrac{19 c_{1,0}}{90\pi} - \tfrac{c_{1,1}}{6\pi}
\right)
\Biggr] + \mathcal{O}(x^6)\,.
\end{align}

\paragraph{Magnitude for ${\ell=2}$:}
\begin{align}
|\mathcal{S}_{\ell=2}| &= 1
+ x\, \pi
+ x^{2} \Bigl[ -\tfrac{79 L}{105}+\tfrac{\pi ^2}{6}+\tfrac{25559}{3675} \Bigr]
+ x^{3} \Bigl[ -\tfrac{79 \pi  L}{105}-\tfrac{\pi ^3}{6}+\tfrac{25559 \pi }{3675} \Bigr] \nonumber\\
&\quad + x^{4} \Bigl[
 \tfrac{4222226168}{121550625} + \tfrac{6241 L^2}{22050} + \tfrac{1453 \pi ^2}{2450}
 - \tfrac{\pi ^4}{40} + \bigl(-\tfrac{1353146}{231525}-\tfrac{79 \pi ^2}{630}\bigr) L - \tfrac{632 \zeta(3)}{105}
\Bigr] \nonumber\\
&\quad + x^{5} \Bigl[
 \tfrac{6241 \pi  L^2}{22050} - \tfrac{38041 \pi ^3}{22050} + \tfrac{19 \pi ^5}{360}
 + \bigl(\tfrac{79 \pi ^3}{630}-\tfrac{1353146 \pi }{231525}\bigr) L
 + \pi  \bigl(\tfrac{4222226168}{121550625}-\tfrac{632 \zeta(3)}{105}\bigr)
\Bigr] + \mathcal{O}(x^6)\,.
\label{eq:S2mag}
\end{align}

\paragraph{Phase for ${\ell=2}$:}
\begin{align}
\tx{Arg} \, \mathcal{S}_{\ell = 2} &= x \bigl[ L_{\tx{IR}}+2 \gamma_E -4 \bigr]
+ x^{2}\, \tfrac{79 \pi }{105}
+ x^{3} \Bigl[ \tfrac{158 \pi ^2}{315}-\tfrac{8 \zeta(3)}{3} \Bigr]
+ x^{4}\, \tfrac{708247 \pi }{1157625} \nonumber\\
&\quad + x^{5} \Bigl[
 -\tfrac{16903}{33075} + \tfrac{1416494 \pi ^2}{3472875} - \tfrac{632 \pi ^4}{4725}
 + \tfrac{49928 \zeta(3)}{11025} + \tfrac{32 \zeta(5)}{5} + \tfrac{c_{2,0}}{60\pi}
\Bigr] + \mathcal{O}(x^6)\,.
\label{eq:S2arg}
\end{align}
The complete expressions through $\mathcal{O}(x^{10})$ are provided in the ancillary file \cite{Sommerfeld_data}.

\section{Resummation}
In this appendix, we give a imposed resummation proposal based on the new RG equation and the perturbative Sommerfeld data. Let us first focus on the amplitude part. We define
\begin{align}
x \equiv GM\omega, \qquad z_1 \equiv \frac{2\omega}{\bar \mu} e^{-\gamma_E} = e^{L/2 - \gamma_E} ~, \quad z_2 \equiv \frac{2\omega}{\bar \mu} ~.
\end{align}
Our proposal separates the amplitude into three parts:
\begin{equation}
    |\mathcal{S}| = |\mathcal{S}|_{\rm IR} \times |\mathcal{S}|_{\rm run} \times |\mathcal{S}|_{\rm rem} ,
\end{equation}
where the first part captures the IR resummation, the second part captures the RG running, and the last part captures the remaining corrections which can be order by order computed by comparing with the perturbative data. The $|\mathcal{S}|_{\rm IR}$ factor contains the Gamma functions and the non-analytic term $(\omega r_s)^{-i\omega r_s}$ in the $n=0$ BHPT wavefunction Eq.~\eqref{eq: Rc def}. This gives
\begin{equation}
    |\mathcal{S}|_{\rm IR} = \left|\frac{\Gamma(\nu(\omega)+1+2ix)\Gamma(2\ell+2)}{\Gamma(2\nu(\omega)+2)\Gamma(\ell+1)}\right| e^{\pi x} .
\end{equation}
Our complete solution of the renormalization group equation also suggests a running factor $|\mathcal{S}|_{\rm run}$
\begin{equation}
\label{eq:resum_run}
    |\mathcal{S}|_{\rm run} = \left|\frac{\bar{F}_{\ell,-}(\omega)-\bar{F}_{\ell,+}(\omega)}{\bar{F}_{\ell,-}(\omega)-\bar{F}_{\ell,0}(\omega)+\bigl(\bar{F}_{\ell,0}(\omega)-\bar{F}_{\ell,+}(\omega)\bigr) z_2^{2(\nu(\omega)-\ell)}}\right| z_1^{(\nu(\omega)-\ell)}.
\end{equation}
To better understand the appearance of this factor, instead of choosing $\mu=\mu_0$ in Eq.~\eqref{eqapp:renom_asymp}, we choose $\mu=\omega$. Then we can write the waveform at inifinity as
\begin{equation}
\begin{aligned}
    R_L |_{r \rightarrow \infty} & = - \frac{c_\ell}{W_{21}(\omega) + c_\ell \bar F(\omega,\mu=\omega,\mu_0) W_{22}(\omega)} \frac{\bar Q_{L}(\omega,\mu=\omega,\mu_0)}{2 i \omega} \frac{e^{i\omega r}}{r}  \\
    & = \frac{\bar{F}_{\ell,-}(\omega)-\bar{F}_{\ell,+}(\omega)}{\bar{F}_{\ell,-}(\omega)-\bar{F}_{\ell,0}(\omega)+\bigl(\bar{F}_{\ell,0}(\omega)-\bar{F}_{\ell,+}(\omega)\bigr)\left( \frac{\omega}{\mu_0} \right)^{2(\nu(\omega)-\ell)}}
\left( \frac{\omega}{\mu_0} \right)^{(\nu(\omega)-\ell)} \\
& \quad \times \Bigg[- \frac{c_\ell}{W_{21}(\omega) + c_\ell \bar F(\omega,\mu=\omega,\mu_0) W_{22}(\omega)} \frac{\bar Q_{L,0}(\omega)}{2 i \omega} \frac{e^{i\omega r}}{r}\Bigg] ~.
\end{aligned}
\label{eq:full_R}
\end{equation}
Eq.~\eqref{eq:resum_run} then appears as a common prefactor. However, we also notice that Eq.~\eqref{eq:resum_run} fails to resum all the $\log(\omega/\mu_0)$ dependence in~Eq.~\eqref{eq:full_R} as there are extra contributions coming from $\bar F(\omega, \mu=\omega,\mu_0)$. These contributions can be computed order by order perturbatively and we decide to put them into the remaining piece.
 
We explicitly show the remaining part for $\ell=1,2$ here until $\mathcal{O}(G^6)$. The complete expressions for $\ell=0,1,2$ through $\mathcal{O}(G^{10})$ are provided in \cite{Sommerfeld_data}
\begin{equation}
\begin{aligned}
    |\mathcal{S}|_{{\rm rem},\ell=1} & = 1  -\frac{223}{450} x^2 + \Big(\frac{7 c_{1,0}}{18 \pi }-\frac{c_{1,1}}{12 \pi }-\frac{629867473}{125685000} \Big) x^4 +\Big(-\frac{19 c_{1,0}}{90}-\frac{c_{1,1}}{6}-\frac{4 \pi }{9}\Big) x^5 \\
    & \quad + \Bigg[L \left(-\frac{133 c_{1,0}}{135 \pi }+\frac{19 c_{1,1}}{90 \pi }-\frac{56}{27}\right)+\zeta_3 \left(-\frac{4 c_{1,0}}{3 \pi }-\frac{160}{57}\right)-\frac{c_{1,0}^2}{288 \pi
   ^2}-\frac{19}{54} \pi  c_{1,0}+\frac{629611 c_{1,0}}{81000 \pi }-\frac{1}{9} \pi  c_{1,1} \\
   & \quad -\frac{7211 c_{1,1}}{5400 \pi }-\frac{7 c_{1,2}}{18 \pi }+\frac{c_{1,3}}{12 \pi
   }-\frac{20 \pi ^2}{27}-\frac{8494493767148059}{428768093250000}\Bigg] x^6 + \mathcal{O}(x^7) ~, \\
    |\mathcal{S}|_{{\rm rem},\ell=2} & = 1  -\frac{3523}{22050}x^2 -\frac{33224323}{36015000} x^4 + \Big(\frac{91 c_{2,0}}{900 \pi }-\frac{c_{2,1}}{60 \pi }-\frac{519376841483437477}{204959587659750000}\Big) x^6 + \mathcal{O}(x^7) ~.
\end{aligned}
\end{equation}

\end{document}